\documentclass{caosp310}

\usepackage{graphicx}
\usepackage{float}
\usepackage{color}

\usepackage{natbib}
\articleNo{}
\pubyear{}
\volume{}
\volnumber{}
\firstpage{}
\received{}
\accepted{}

\def\BibTeX{{\rm B\kern-.05em{\sc i\kern-.025em b}\kern-.08em
             T\kern-.1667em\lower.7ex\hbox{E}\kern-.125emX}}

\begin{document}

%
\htitle{Aluminium abundances of B-type stars}
\hauthor{Y. Takeda}

\title{Formation of Al\,{\sc ii} lines and photospheric aluminium abundances 
in B-type stars}


%
%
\author{
        Yoichi Takeda\orcid{0000-0002-7363-0447}
       }

%
\institute{
          11-2 Enomachi, Naka-ku, Hiroshima-shi, 730-0851, Japan \\
          \email{ytakeda@js2.so-net.ne.jp}
          }

\date{}

\maketitle

\begin{abstract}
The aluminium abundances of early-to-late B-type main-sequence stars in 
the effective temperature range of 10000\,K\,$\la T_{\rm eff} \la$\,22000\,K 
(comprising normal stars as well as chemically peculiar HgMn stars) 
were spectroscopically determined, with an aim of getting information 
about the galactic gas composition at the time of their formation 
from their photospheric abundances.
For this purpose, two Al\,{\sc ii} lines at 6243 and 4663\,\AA\ were employed, 
for which the non-LTE effect was taken into account based on detailed 
statistical-equilibrium calculations. 
The non-LTE effect of these Al\,{\sc ii} lines generally acts in the 
direction of weakening (i.e., profile becomes shallower) caused by a 
decrease of line opacity (due to overionization) along with an enhanced 
line source function (overexcitation), and this effect tends to become 
progressively larger with an increase in $T_{\rm eff}$ as well as 
with a decrease in $\log g$ (surface gravity).
Regarding the Al\,{\sc ii} 6243 line, while the non-LTE calculation 
qualitatively reproduces its overall behavior (e.g., transition from 
absorption to emission at early B-type), some $T_{\rm eff}$-dependent
systematic trend remains unremoved in the non-LTE abundances of normal 
stars, which means that non-LTE corrections evaluated for this line 
are quantitatively insufficient. Meanwhile, for the case of the 
Al\,{\sc ii} 4663 line, which is more advantageous than the 6243 line 
in the sense that it is stronger without showing any emission,
the resulting non-LTE abundances of ordinary B stars are almost 
constant at the solar abundance ($A \simeq 6.5$) over the wide 
$T_{\rm eff}$ range ($\sim$\,10000--20000\,K), suggesting that the abundances 
derived from this line are successfully non-LTE-corrected and trustable.
Therefore, according to the results from the Al\,{\sc ii} 4663 line, 
we may conclude that the Al abundance of the galactic gas in 
the recent past (several times $\sim 10^{7}$--$10^{8}$\,yr ago)
is almost consistent with the solar composition.
As to the Al abundances of HgMn stars ($T_{\rm eff} \la 15000$\,K), 
our analysis confirmed that this element is conspicuously deficient 
(by $\sim$\,0.5--2\,dex in comparison with the Sun) in 
the photosphere of these chemically peculiar stars, as already reported 
in previous studies. 
\end{abstract}

\section{Introduction}

Aluminium (Al; $Z=13$) is one of the intermediate elements of astrophysical 
importance, which is considered to be synthesized mainly during the hydrostatic 
carbon and neon burning within massive stars and expelled outwards by type~II 
supernovae explosion. While its chemical abundances have been rather well 
determined for a number of late-type (F--G--K) stars in a wide range of metallicity
(e.g., Baum\"{u}ler and Gehren, 1997; Andrievsky {\it et al.}, 2008; and the references 
therein), the situation is still insufficient in the field of early-type (A--B) 
stars. Unlike the case of cool solar-type stars where lines common to solar abundance 
determination may be used for deriving [Al/H] (differential abundance relative 
to the Sun), what matters more seriously in the case of hot stars is the 
reliability of absolute abundances which are not easy to establish precisely. 
Especially, since most of the previous studies have been done under the 
conventional assumption of LTE (Local Thermodynamic Equilibrium), an inadequacy 
of this presumption (neglect of the non-LTE effect) may be counted as a possible 
source of systematic error.

Given this situation, Takeda (2023; hereinafter referred to as Paper~I) tried 
to determine Al abundances of A-type stars (7000\,K\,$\la T_{\rm eff} \la$\,10000\,K)
based on the Al\,{\sc i} 3944/3961 resonance lines by taking into account
the non-LTE effect, because LTE abundances derived by past investigators were 
suspected to be considerably underestimated.
It revealed in Paper~I that this doublet suffers an appreciable non-LTE 
line-weakening and the serious zero-point discrepancy could be successfully removed
([Al/H] $\sim 0$ for [Fe/H]\,$\sim 0$ stars) by applying the significant (positive) 
non-LTE corrections amounting to up to $\la 1$\,dex.

As the follow-up of Paper~I, the present study focuses on B-type stars
(10000\,K\,$\la T_{\rm eff} \la$\,22000\,K), but the intended scope  
is somewhat different. Most comparatively sharp-lined A-type stars 
(for which spectroscopic abundance determinations are feasible) more or less 
show some kind of chemical peculiarity (CP), and thus ``normal'' A stars are 
difficult to find in practice. In contrast, B-type main-sequence 
stars are divided rather clearly into non-CP (normal) stars and CP stars (mainly 
HgMn stars of late B-type), and we may expect the former ``ordinary B stars''  
to retain the original gas at the time of star formation. 
Therefore, it may be possible to get the compositional information of 
the galactic gas in the recent past ($\sim 10^{7-8}$\,yr ago 
when these B-type stars of $\sim$\,3--9\,$M_{\odot}$ were formed) from the
the photospheric abundances of normal B stars.

However, Al abundance determinations of B-type stars seem to have been more
focused on chemically peculiar HgMn stars because of their anomalous aspects, 
for which considerable Al-deficiency has been almost established (see the compilation 
of Ghazaryan and Alecian, 2016).
Meanwhile, our understanding on the abundances of this element for normal B stars 
in comparison with the Sun (are they consistent with the solar abundance?) is still 
unsatisfactory, because published results 
(all derived with the assumption of LTE) 
tend to be diversified depending upon 
the adopted lines (UV or optical; Al\,{\sc i} or Al\,{\sc ii} or Al\,{\sc iii}), 
as summarized below (though this literature survey may not be complete).
\begin{itemize}
\item[$\bullet$]
Al abundance determinations for the well-studied benchmark sharp-lined 
star $\iota$~Her (B3\,IV) have been reported by several authors, 
as compiled by Golriz and Landstreet (2017; cf. Table~1 therein). 
While most of them are around $A$\footnote{$A$ is the logarithmic 
number abundance of the element relative to that of hydrogen (H) with the 
usual normalization of $A$(H) = 12.}$\sim 6.5$ and consistent with the solar 
abundance ($A_{\odot} = 6.47$),\footnote{In this article, Anders and Grevesse's 
(1989) solar photospheric Al abundance of $A_{\odot} = 6.47$ is adopted as the 
reference, in order to keep consistency with Kurucz's (1993) ATLAS9/WIDTH9 
program. This is fairly close to values given in Asplund {\it et al.}'s (2009) 
Table~1 ($6.45\pm0.03$ for the solar photosphere, $6.43\pm0.01$ for 
the meteorites).} Pintado and Adelman (1993) derived an appreciably discrepant 
result ($A = 5.97$) from the Al\,{\sc ii} 4663 line (though they obtained 
a near-solar averaged abundance of $\langle A \rangle \sim 6.5$ from four 
Al\,{\sc iii} lines).
\item[$\bullet$]
Pintado and Adelman (1993) also determined the Al abundance of $\gamma$~Peg 
(B2\,IV) and found that both Al\,{\sc ii} and Al\,{\sc iii} lines yielded a 
consistent abundance of $\langle A \rangle \sim$\,6.0--6.1 (subsolar by several 
tenths dex). 
\item[$\bullet$]
In Allen's (1998) abundance studies on early A and late B stars, attempts 
of Al abundance determinations for normal late B-type stars (21~Peg, $\zeta$~Dra,
21~Aql, $\tau$~Her) based on Al\,{\sc i} 3944/3961 and Al\,{\sc ii} 4663 lines 
yielded $A \sim $\,6.3--6.4 (almost solar or slightly subsolar), while five HgMn stars 
were confirmed to be considerably Al-deficient (by $\sim 1$\,dex or even more). 
\item[$\bullet$]
Fossati {\it et al.} (2009) determined Al abundances of two normal late B-type stars 
(21~Peg and $\pi$~Cet) and obtained [Al/H]\footnote{As usual, [X/H] is the 
differential abundance of element X relative to the Sun; i.e., 
[X/H]\,$\equiv A_{\rm star}({\rm X})- A_{\odot}({\rm X})$.}\,$\sim 0$\,dex 
(almost solar) for both, though the abundance from Al\,{\sc iii} lines (measured 
only for $\pi$~Cet) turned out appreciably higher by $\sim 0.4$\,dex. 
\item[$\bullet$]
Niemczura {\it et al.} (2009) conducted an extensive abundance study on a number of 
late B-type stars (including chemically peculiar stars). Although distinction of 
normal and peculiar stars is not clear in their sample and the adopted lines 
are not explicitly described, the Al abundances they obtained range widely from 
$A \sim 5.1$ to $\sim 6.8$.    
\item[$\bullet$]
The Al abundances of the B-type supergiant $\beta$~Ori (B8\,Iae) derived by 
Przybilla {\it et al.} (2006) show an appreciable discrepancy between Al\,{\sc ii} 
($6.14 \pm 0.08$) and Al\,{\sc iii} ($7.00 \pm 0.38$) lines. 
\item[$\bullet$]
Monier (2022) carried out an Al abundance analysis for HD~209459 (whether it is
normal or peculiar is still controversial) based on the Al\,{\sc ii} 4663 line 
in the optical region and the Al\,{\sc iii} 1854/1862 resonance lines in the 
ultraviolet region. He found that, while the the strength of the former optical 
line is consistent with the solar Al abundance, 
a significantly reduced Al abundance ($\sim$\,30\% of the solar composition)
is needed to reproduce the strength of the latter UV lines.   
\end{itemize}

Recently, Takeda (2024; hereinafter referred to as Paper~II) determined the phosphorus
abundances of B-type stars by using the P\,{\sc ii} 6043 line while taking into account 
the non-LTE effect, and found that the non-LTE P abundances of superficially normal 
early-to-late B-type stars (formed several times $\sim 10^{7}$--$10^{8}$\,yr ago) are 
systematically lower than the abundance of the Sun (its age is $\sim 4.6 \times 10^{9}$\,yr) 
by $\sim$\,0.2--0.3\,dex. This is a significant conclusion, because it means that 
the galactic gas composition of P has decreased with time, in contradiction to 
the general concept of chemical evolution.

It is, therefore, worthwhile to examine how the abundances of Al (intermediate 
element similar to P) in normal B-type stars are compared with that of the Sun   
(near-solar as usually expected? or subsolar like the case of P?), since
published determinations are too insufficient to answer this question
as mentioned above.

This situation motivated the author to contend with the task of Al abundance
determinations for the same sample of B-type stars as investigated in Paper~II.
This is the purpose of the present study. Towards this aim, special attention 
is paid to the following points in context of the unsatisfactory results of 
past publications.
\begin{itemize}
\item
One of the reasons for the confusingly diversified literature results may be 
the mixed use of various kinds of Al lines for abundance determination. 
In order to make things simple, we restrict ourselves in this investigation 
to two Al\,{\sc ii} lines in the optical region at 6243 and 4663\,\AA, 
both of which are observable from early- through late-B stars of wide 
$T_{\rm eff}$ range and of sufficiently high quality 
(e.g., neither too strong nor too weak, almost free from blending).   
\item
The non-LTE effect is taken into account in deriving the abundances from these 
Al\,{\sc ii} lines. This may be significant, because previous Al abundance 
determinations for B stars mentioned above were done under the assumption of LTE.
Of course, there is no guarantee that computed non-LTE corrections are reliable, 
since statistical-equilibrium calculations may often be imperfect due to uncertainties 
in the adopted atomic model. Yet, the validity of the non-LTE calculation can 
be checked by examining whether the resulting non-LTE Al abundances of normal 
B-type stars do not show any systematic dependence upon $T_{\rm eff}$ (most 
important parameter), which actually makes a useful touchstone.
\end{itemize}

\section{Line-formation of Al\,{\sc ii} lines}

\subsection{Non-LTE calculations}

Statistical-equilibrium calculations for Al\,{\sc ii} were carried out
by using the model atom of 68 terms (up to 10$p$\,$^{3}$P$^{\rm o}$ at 
146602\,cm$^{-1}$) with 601 radiative transitions (cf. Sect.~2.1 in Paper~I
for more details). 
While the contribution of Al\,{\sc i} was neglected
because it is negligible in the atmosphere of B-type stars, that of Al\,{\sc iii} 
(32 terms) and Al\,{\sc iv} (only the ground term) was taken into account 
in the number conservation of total Al atoms. 
As done in Paper~I, the data of photoionization cross sections were taken from
TOPbase (Cunto and Mendoza, 1992) for the lowest 10 Al~{\sc ii} terms (while the 
hydrogenic approximation was applied for the remaining terms), and the collisional
rates were evaluated by following the recipe described in Sect.~3.1.3 of Takeda (1991).  

The non-LTE departure coefficients were calculated on a grid of 56 
($= 14 \times 4$) solar-metallicity model atmospheres resulting from 
combinations of fourteen $T_{\rm eff}$ values (9000, 10000, 11000, 12000, 
13000, 14000, 15000, 16000, 17000, 18000, 19000, 20000, 22000, and 24000\,K) 
and four $\log g$ values (3.0, 3.5, 4.0, and 4.5).while assuming 
$v_{\rm t}$ = 2\,km\,s$^{-1}$ (microturbulence) and the solar abundance 
of $A$(Al) = 6.47 ([Al/H] = 0) as the input Al abundance.\footnote{The 
effect of assigned Al abundance on departure coefficients is discussed
in Sect.~5.1.} 

\subsection{Theoretical profiles and strengths of Al\,{\sc ii} lines}

Now that the departure coefficients have been computed, the profile and 
strength of any line can be computed. Here, attention is paid to three 
representative Al\,{\sc ii} lines at 6243, 4663, and 3900\,\AA.
The former two (6243 and 4663) are the lines used for abundance determination,
while the latter 3900 line was originally considered as 
a candidate of abundance indicator but eventually abandoned (cf. Sect.~5.2.3).
The atomic data of these lines are summarized in Table~1. 

Fig.~1 illustrates how the theoretical equivalent widths of these lines calculated 
in LTE ($W^{\rm L}$; dashed line) as well as in non-LTE ($W^{\rm N}$; solid line) 
and the non-LTE corrections ($\Delta \equiv A^{\rm N} - A^{\rm L}$, 
where $A^{\rm L}$ and $A^{\rm N}$ are the abundances derived from $W^{\rm N}$ 
with LTE and non-LTE) depend upon $T_{\rm eff}$ and $\log g$. Likewise, 
the non-LTE and LTE profiles of these lines for the models of different 
$T_{\rm eff}$ values (10000, 14000, 18000, and 22000\,K) are shown in Fig.~2.

The following characteristics are observed from these figures.
\begin{itemize}
\item[$\bullet$]
The non-LTE effect acts as a line-weakening mechanism ($W^{\rm N} < W^{\rm L}$;
Fig.1 and 2), which results in positive non-LTE corrections ($\Delta > 0$). 
\item[$\bullet$]
The non-LTE corrections ($\Delta$) tend to increase towards higher $T_{\rm eff}$
and with a decrease in $\log g$ (Fig.~1). 
\item[$\bullet$]
The overall behavior of $W$ is qualitatively rather similar to each other 
with a peak around $T_{\rm eff} \sim$\,12000--14000\,K (though quantitatively 
$W_{6243} < W_{4663} < W_{3900}$). 
\item[$\bullet$]
However, an emission feature emerges in the non-LTE profile of Al\,{\sc ii} 6243 
line at $T_{\rm eff} \ga 17000$\,K (leading to $W^{\rm N}_{6243} < 0$),
despite that the other two 4663 and 3900 lines retain ordinary absorption 
profiles. 
\end{itemize}

\subsection{Physical mechanism of the non-LTE effect}

In order to understand the behaviors of the non-LTE effect described 
in Sect.~2.2, the non-LTE-to-LTE line-center opacity ratio 
$l_{0}^{\rm NLTE}/l_{0}^{\rm LTE}$
($\simeq b_{\rm l}$) and the ratio of the line source function to 
the Planck function $S_{\rm L}/B$ 
($\simeq b_{\rm u}/b_{\rm l}$,\footnote{
This relation holds under the condition that the photon energy $h\nu$ 
($h$: Planck constant, $\nu$: frequency) is not small in comparison 
with $kT$ ($k$: Boltzmann constant). See also footnote~6.} where $b_{\rm l}$ 
and $b_{\rm u}$ are the departure coefficients of lower and upper levels) 
for the transitions relevant to Al\,{\sc ii} 6243, 4663, and 3900 lines are 
plotted against the optical depth in  Fig.~3, 4, and 5, respectively. 

We can see from these figures that
$l_{0}^{\rm NLTE}/l_{0}^{\rm LTE} < 1$ (overionization) and
$S_{\rm L}/B > 1$ (overexcitation) in the line-forming region.
This explains why the strengths of these lines are generally decreased 
by the non-LTE effect in B-type stars ($W^{\rm N} < W^{\rm L}$), because 
both conditions act in the direction of weakening the absorption profile.   

Especially, the latter effect ($S_{\rm L}/B > 1$) plays a comparatively more 
significant role in this line weakening, which becomes more conspicuous 
with an increase in $T_{\rm eff}$. Actually, the value of $S_{\rm L}/B $ 
in the line-forming region (e.g., at $\tau_{5000} \sim 0.1$) tends to be 
progressively larger towards higher $T_{\rm eff}$ (see panels 
(f)$\rightarrow$(g)$\rightarrow$(h) in Fig.~3 or Fig.~4), which eventually 
results in the trend of increasing $\Delta$ with $T_{\rm eff}$.    

The cause of such an overexcitation ($S_{\rm L}/B > 1$) may be interpreted 
as that the lower level is more overionized than the upper level. 
Generally, overionization takes place when the photoionization rate 
($\propto J_{\lambda} \sim W_{\rm D} B_{\lambda}(T_{\rm R})$) outweighs
the recombination rate ($\propto B_{\lambda}(T_{\rm e})$) as
$W_{\rm D} B_{\lambda}(T_{\rm R}) > B_{\lambda}(T_{\rm e})$, where 
$T_{\rm R}$ is the radiation temperature ($\simeq$ temperature of the
continuum forming region), $W_{\rm D}$ is the dilution factor, and 
$T_{\rm e}$ is the electron temperature of the line-forming region.
Since this inequality critically depends upon the $T$-sensitivity of
$B_{\lambda}$, overionization is more enhanced at shorter wavelength
regions (where $B_{\lambda}$ is more sensitive to $T$). 
Accordingly, the fact that the wavelength of the photoionization edge for 
the lower level of a line is shorter than that for the upper level
leads to $b_{\rm l} < b_{\rm u} (< 1)$; i.e., $S_{\rm L} > B$.

The appearance of an emission feature in the Al\,{\sc ii} 6243 line at 
the high $T_{\rm eff}$ range ($\ga 17000$\,K) resulting from our non-LTE 
calculation (Fig.~2c and 2d) is also due to this effect of appreciably 
larger $S_{\rm L}$ over $B$, as recognized from Fig.~3g and 3h. 
The reason why this line shows an emission in the regime of early B-type 
may presumably be related to its longer wavelength than the other two, 
because the departure of $S_{\rm L}$ from $B$ tends to be further enhanced 
in the Rayleigh--Jeans region ($h\nu/kT < 1$).\footnote{
Let us define a parameter $\beta$ $(\equiv 1- b_{\rm l}/b_{\rm u})$ in order to
indicate the degree of non-LTE overexcitation (i.e., difference between
$b_{\rm l}$ and $b_{\rm u}$), by which $S_{\rm L}/B$ is
expressed as $\simeq 1/(1-\beta)$ in the often encountered case of $h\nu/kT > 1$ 
(cf. footnote~5).  However, in the high $T$ and low $\nu$ case of $h\nu/kT < 1$,
an alternative relation $S_{\rm L}/B \simeq 1/(1-\beta/\delta)$ holds,
where $\delta \equiv h\nu/kT (< 1)$. That is, the extent of non-LTE departure 
($\beta$) is further exaggerated by a factor of $1/\delta (>1)$. 
See Sect. 12-4 in Mihalas (1978) for more details.} It is worth noting that 
the Si\,{\sc ii} 6239.6 line (which is a similar high-excitation line with
$\chi_{\rm low} \simeq 13$\,eV like Al\,{\sc ii} 6243; cf. Table~1) also
shows an emission in early-to-mid B-type stars (see Sect.~4.3 and Fig.~6).

\section{Observational data}

\subsection{Program stars}

The target stars adopted in this investigation for Al abundance determinations 
are the same as in Paper~II, which are 85 early-to-late B-type  
(near-)main-sequence stars (comprising normal as well as HgMn peculiar stars) 
in the solar neighborhood. They are comparatively sharp-lined 
($v_{\rm e}\sin i \la 60$\,km\,s$^{-1}$) and have masses in the range of 
$2.5 M_{\odot} \la M \la 9 M_{\odot}$. The list of these 85 program 
stars is given in Table~2.

\subsection{Adopted spectra}

Since the sample stars in Paper~II are targeted unchanged in this study, 
it is natural to use the same observational data (high-dispersion
spectra obtained at Okayama Astrophysical Observatory) as adopted therein 
(see Sect.~2 in Paper~II for the details). However, since the wavelength range 
covered by these spectra are limited to visible--red (or near-IR) region 
($\lambda \ga 5000$\,\AA), they can be employed only for the analysis of 
Al\,{\sc ii} 6243 line in the orange region.

Accordingly, blue-region spectra necessary for the analysis of Al\,{\sc ii} 4663 line 
(or Al\,{\sc ii} 3900 line) had to be obtained from the public-open data of stellar 
spectra. After searching the databases of ESO, CFHT, and ELODIE, spectra of 51 stars 
(out of 85 program stars) were found to be available and thus downloaded. The details 
of these data (file names, etc.) are summarized in Table~3.

\section{Abundance determination}

\subsection{Procedures}

The abundances of Al for the program stars are determined by almost the 
same procedures as adopted in Paper~I or Paper~II, which consist of four steps. 
(1) First, the spectrum-fitting analysis is applied to two spectral regions 
comprising Al\,{\sc ii} 6243 and Al\,{\sc ii} 4663 lines, 
and the best fit parameter solutions are determined. 
(2) Based on such established abundance solutions, the equivalent widths 
of these two Al lines ($W$) are inversely calculated. (3) Then the non-LTE 
abundance ($A^{\rm N}$) is derived from $W$ by taking into account the non-LTE 
effect. (4) Finally, uncertainties in the abundance results are estimated, 
while considering the errors in $W$ along with the sensitivities to parameter 
changes. 

\subsection{Atmospheric parameters}

Regarding the atmospheric parameters assigned to each star, the same values 
as used in Paper~II (see Sect.~3 therein for the details) are adopted 
in this study. In Table~1 are shown the values of $T_{\rm eff}$ and $\log g$, 
which were originally determined from colors of the Str\"{o}mgren system.  
As to the microturbulence, $v_{\rm t}$ = 1\,km\,s$^{-1}$ ($T_{\rm eff} < 16500$\,K) and  
2\,km\,s$^{-1}$ ($T_{\rm eff} > 16500$\,K) are tentatively assumed as done in Paper~II,
which is sufficient because  this parameter is practically insignificant in the 
present case (cf. Sect.~4.5). Likewise, the model atmospheres for each of 
the targets are the solar-metallicity models generated by interpolating 
Kurucz's (1993) ATLAS9 model grid in terms of $T_{\rm eff}$ and $\log g$.

\subsection{Synthetic spectrum fitting}

The spectrum-fitting analysis (Takeda, 1995) by assuming LTE was applied to 
two wavelength regions (6230--6250\,\AA\ and 4656--4668\,\AA, each comprising 
Al\,{\sc ii} 6243 and 4663, respectively), in order to accomplish the best fit 
between theoretical and observed spectra. Here, the parameters varied are 
the chemical abundances (of Al and other elements showing appreciable lines), 
rotational broadening velocity ($v_{\rm e}\sin i$), and radial velocity 
($V_{\rm rad}$). The data of all atomic lines (including those of Al lines
given in Table~1) in each region were taken from the VALD database 
(Ryabchikova {\it et al.}, 2015). 

It then revealed that, while Al abundances could be successfully established for 
all 51 stars in the fitting analysis of the 4656--4668\,\AA\ region, its determination 
had to be abandoned for 29 stars (out of 85 stars) in the 6230--6250\,\AA\ region analysis  
since the Al\,{\sc ii} 6243 line did not exhibit any detectable absorption profile.
This is due to the fact that this line begins to show an emission feature in the high 
$T_{\rm eff}$ regime  of early-to-mid B-type stars, which is actually predicted
in the non-LTE calculations (cf. Fig.~2c and 2d). In order to illustrate this
situation, the spectral portions in the neighborhood of the Al\,{\sc ii} 6243 and 
4663 lines for stars with $T_{\rm eff} \ga 16000$\,K are displayed in Fig.~6, 
where we can see that the Al\,{\sc ii} 6243 line is not usable for a large fraction 
of stars at $T_{\rm eff} \ga 17000$\,K (because of being emission or very weak due to
filled-in emission). Meanwhile, this 6243 line is too weak to be measurable 
in chemically peculiar HgMn stars (mostly late B-type) because Al is 
considerably deficient in their atmospheres.

Accordingly, those 29 stars for which Al abundance could not be derived from 
the 6243 line are divided into three categories: (i) emission line  
(early- or mid-B stars), (ii) very weak line due to the effect of 
filled-in emission (mid-B stars), and (iii) very weak line due to 
considerable Al-deficiency (late-B HgMn stars), as indicated in 
column 13 of Table~2. In such cases, the fitting was tentatively done either 
by masking the 6243 line region or by fixing the Al abundance at an arbitrary value. 
The cases (ii) and (iii) can be clearly distinguished from each other
by checking the neighboring Si\,{\sc ii} 6239.6 line, because this Si\,{\sc ii} line
also shows an emission (or filled-in emission) in the former case (ii), while 
not in the latter case (iii).

The comparison of theoretical spectrum (for the solutions 
with converged parameters) with the observed spectrum
(for the selected wavelength region in the neighborhood of the relevant Al lines) 
is shown in Fig.~7 (6243 line region) and Fig.~8 (4663 line region) for each star.

\subsection{Evaluation of equivalent widths}

Next, the equivalent widths ($W_{6243}$ and $W_{4663}$) of the Al\,{\sc ii} 6243 
and 4663 lines were inversely evaluated from the Al abundance solution 
derived from the spectrum-fitting analysis (cf. Sect.~4.3) with the same model 
and atmospheric parameters, where Kurucz's (1993) WIDTH9 program\footnote{
The original WIDTH9 program was considerably modified by the author in various respects;
for example, enabling the treatment of multi-component lines (such as the Al\,{\sc ii} 
6243 line), inclusion of the non-LTE effect, etc.}  
was employed for this purpose. 
The errors involved in such obtained $W$ values ($\delta W$) were further estimated 
from the line-center depth and the S/N ratio by applying Eq.~(1) in Paper~II,
which are typically on the order of $\la$\,1--2\,m\AA\ in most cases 
(though up to several m\AA\ or more in exceptionally shallow/broad-line cases) 

The resulting $W_{6243}$ and $W_{4663}$ for each star are presented in Table~2.
Regarding 29 stars for which abundances could not be determined from the 6243 line
(cf. Sect.~4.3), following values are assigned to $W_{6243}$ in this table
for each of the three cases (i, ii, iii).
(i) $W_{6243}$ (negative) was directly measured from the emission-line profile
by the Gaussian fitting. (ii) Zero value (0) is tentatively given.
(iii) An upper-limit value ($W_{6243}^{\rm ul}$) is presented, which was estimated 
from the line width and the S/N ratio by using Eq.~(1) of Takeda (2025).

\subsection{Non-LTE abundances and their uncertainties}

Then, the non-LTE abundances ($A^{\rm N}_{6243}$ and $A^{\rm N}_{4663}$) were 
determined from  $W_{6243}$ and $W_{4663}$ for each star by taking into account 
the departure from LTE, where the departure coefficients calculated on the grid 
of 56 models (cf. Sect.~2.1) were interpolated in terms of $T_{\rm eff}$ and $\log g$, 
from which the non-LTE corrections ($\Delta$; difference of $A^{\rm N}$ from 
the LTE abundance $A^{\rm L}$) were also obtained.
The resulting values of $W$, $A^{\rm N}$, and $\Delta$ for each line are given 
in Table~2. Likewise, $W$, $A^{\rm L}$, $\Delta$ and $A^{\rm N}$ are plotted against 
$T_{\rm eff}$ in panels ((a)--(d)) of Fig.~9 (6243) and Fig.~10 (4663),    
where the error bars attached in $W$ (panel (a)) are $\pm \delta W$ (Sect.~4.4)

The abundance sensitivities to typical ambiguities in atmospheric parameters 
[$\delta_{T\pm}$ (abundance changes for $T_{\rm eff}$ perturbations by $\pm 3$\%), 
$\delta_{g\pm}$ (abundance changes for $\log g$ perturbations by $\pm 0.2$\,dex), 
and $\delta_{v\pm}$ (abundance changes for $v_{\rm t}$ perturbations by $\pm 1$\,km\,s$^{-1}$)] 
are plotted against $T_{\rm eff}$ in panels (e)--(g) of Fig.~9 and Fig.~10. 
We can see from these figures that combined impacts of parameter uncertainties upon the 
abundances (due mainly to $T_{\rm eff}$ and partly to $\log g$, while the response to  
$v_{\rm t}$ change is negligibly small) are $\pm \la$\,0.1--0.2\,dex
at most. The error bars attached in $A^{\rm N}$ (panel~(d)) are the root-sum-square of
$\delta A$ (abundance error due to $W$ perturbation of $\pm \delta W$), $\delta_{T}$, 
$\delta_{g}$, and $\delta_{v}$.   

\section{Discussion}

\subsection{Impact of assigned abundance in non-LTE calculations}

Before discussing the results obtained in Sect.~4, some comments may be in order 
regarding the Al abundance adopted in non-LTE calculations (Sect.~2.1), 
which we assumed the solar abundance ([Al/H] = 0). 
Since the departure coefficients resulting from statistical-equilibrium calculations
more or less depend upon the input abundance, its validity in an application 
to actual stars needs to be checked.

Originally, it was intended to derive several non-LTE abundances for a star 
corresponding to different sets of departure coefficients calculated with various 
[Al/H] values, and obtain the consistent abundance solution by interpolation
as done in Paper~II (cf. Sect.~6.2 therein).
Unfortunately, this approach did not work well in the present case, 
because $A^{\rm N}_{6243}$ determination (for a given $W_{6243}$) 
was found to be not always successful if [Al/H] is changed.

This situation is illustrated in Fig.~11, where the non-LTE corrections
($\Delta_{0}$, $\Delta_{-0.5}$, $\Delta_{-1}$) for each star derived by applying 
three sets of departure coefficients (corresponding to [Al/H] = 0.0, $-0.5$, 
and $-1.0$) are plotted against $T_{\rm eff}$. An inspection of this figure reveals
that, while $\Delta_{4663}$ is insensitive to a change in [Al/H] (Fig.~11b), 
$\Delta_{6243}$ is appreciably [Al/H]-dependent (i.e., increasing with a decrease 
in [Al/H]; cf. Fig.~11a). More seriously, $\Delta_{6243}$ tends to be indeterminable 
as [Al/H] is decreased, especially at higher $T_{\rm eff}$ (note that  
the number of $\Delta_{6243}$ plotted in Fig.~11a becomes progressively fewer with 
a decrease in [Al/H]). This stems from the fact that the non-LTE line-weakening effect 
(due to the growth of filled-in emission caused by an increase of $S_{\rm L}$) becomes 
more important with a decrease in [Al/H], particularly in the higher $T_{\rm eff}$ regime. 
In such cases, non-LTE $W^{\rm N}_{6243}$ may not necessarily be a monotonically 
increasing function of $A$ any more, since even an increase of the abundance can 
``weaken'' the line strength by the effect of filled-in emission as the line-forming 
region shifts towards higher. Therefore, the solution of $A^{\rm N}_{6243}$ 
corresponding to a given $W_{6243}$ tends to be undetermined at higher $T_{\rm eff}$
if departure coefficients for lower [Al/H] are used. 

However, our choice of applying the non-LTE departure coefficients calculated with 
[Al/H] = 0 to all stars is reasonably sufficient from a practical point of view 
for the following reasons.
\begin{itemize}
\item
Regarding normal B-type stars, since their Al abundances turned out almost solar 
as will be concluded in Sect.~5.3.1, the assumption of [Al/H] = 0 is justified.  
\item
On the other hand, since chemically peculiar HgMn stars are considerably Al-deficient
(cf. Sect.~5.3.2), applying the non-LTE departure coefficients calculated with 
[Al/H] = 0 is not consistent. Nevertheless, because Al abundances of such HgMn stars 
are not established from the problematic 6243 line (which yields only the upper 
limits in most cases), we have to anyhow invoke those ($A^{\rm N}_{4663}$) from 
the 4663 line. Then, since $\Delta_{4663}$ values are insensitive to [Al/H] 
(Fig.~11b), use of the [Al/H] = 0 set even in the 4663 line analysis of Al-deficient 
stars would not cause any serious problem. 
\end{itemize}

\subsection{Characteristics of each Al\,{\sc ii} line}

Based on the results obtained in Sect.~4,  we discuss the abundances derived from 
Al\,{\sc ii} 6243 and 4663 lines and assess their reliability and usability as 
abundance indicator. In addition, the problems involved with the Al\,{\sc ii} 3900 
line, which was eventually abandoned to use, are also described. 
The discussion in this subsection is primarily confined to normal B-type stars, 
which may serve as a touchstone because their prospective abundance homogeneity.

\subsubsection{Al\,{\sc ii} 6243}

The LTE abundances ($A^{\rm L}_{6243}$) derived from this line are considerably 
$T_{\rm eff}$-dependent (progressively decreasing with $T_{\rm eff}$) as shown in
Fig.~9b. Although this trend is surely mitigated by applying the non-LTE corrections
($\Delta_{6243}$; Fig.~9c), some systematic tendency remain unremoved 
in the non-LTE abundances ($A^{\rm N}_{6243}$; Fig.~9d), which means that
they are insufficiently undercorrected. Therefore, our non-LTE calculations for 
this 6243 line are quantitatively still imperfect, despite that they satisfactorily 
predict the qualitative trend (e.g., emergence of emission
feature at higher-$T_{\rm eff}$ regime; see Fig.~ 9a in comparison with  Fig.~1a).
Considering the remarkable [Al/H]-dependence of non-LTE corrections (Fig.~11).
we may regard that the non-LTE results of this 6243 line are vulnerable to inadequacy 
in the adopted conditions (parameters) of the calculations, and thus less reliable 
in the quantitative sense.  

\subsubsection{Al\,{\sc ii} 4663}

The LTE abundances ($A^{\rm L}_{4663}$) determined from the 4663 line also 
show some systematic tendency (decreasing with $T_{\rm eff}$; Fig.~10b). 
However, after the non-LTE corrections ($\Delta_{4663}$; Fig.~10c) have been 
applied, this trend is almost removed in the non-LTE abundances 
($A^{\rm N}_{4663}$) as can be confirmed in Fig.~10d.
The exceptional outliers are two early B-type stars (HD~000886 and HD~035708) 
at $T_{\rm eff} \ga 21000$\,K, which might indicate that the reliability of the 
calculation may deteriorate at such a highest $T_{\rm eff}$ regime.      
The fact that the non-LTE abundances of normal B-type stars derived 
from this 4663 line are almost constant around $A \simeq 6.5$ over a wide
temperature range (10000\,K\,$\la T_{\rm eff} \la 20000$\,K) suggests that
this line is more reliable than the 6243 line as long as Al abundance
determination is concerned. Besides, that this 4663 line is stronger
(abundances are determinable even for Al-deficient stars) and does not 
show any apparent emission (even in early B stars) may be counted as
additional evidence that this line is more advantageous than the 6243 line.  

\subsubsection{Al\,{\sc ii} 3900}

This Al\,{\sc ii} 3900.675 line was originally selected as a potential candidate 
for Al abundance determination, because it was expected to have a sufficient strength 
(even stronger than the 6243/4663 lines) based on preparatory calculations 
by using the oscillator strength taken from VALD ($\log gf = -1.27$). 
However, observed strengths of this line in actual spectra were found to be much 
weaker than this expectation, suggesting the necessity of examining the reliability 
of $\log gf$(VALD) by checking other databases.

It was then found that, while Kurucz and Bell's (1995) compilation presents the 
same value as VALD, the NIST database\footnote{The Atomic Spectra Database of 
the National Institute of Standards and Technology. Available at 
https://www.nist.gov/pml/atomic-spectra-database.
} gives a considerably lower value of $\log gf = -2.26$. 
Accordingly, this $\log gf$(NIST) value was eventually adopted for the Al\,{\sc ii} 3900
line in this study (cf. Table~1), which is regarded as being more reliable than $\log gf$(VALD).
  
Another problem involved with this Al\,{\sc ii} 3900 line is that it is severely 
contaminated by the neighboring Ti\,{\sc ii} line at 3900.539\,\AA, which is even stronger 
than the Al\,{\sc ii} line especially at lower $T_{\rm eff}$, though its strength
drops down at higher $T_{\rm eff}$ (see Fig.~1c).

The situation mentioned above is graphically demonstrated in Fig.~12, where the observed and 
theoretically synthesized spectra around $\sim$\,3900\,\AA\ region are compared with each other 
for HD~160762 (17440\,K), HD~209008 (15353\,K), and HD~209459 (10204\,K), which are normal 
stars with near-solar Al abundances of $A^{\rm N}_{4663} \sim 6.5$ (see Table~2).
At any rate, this Al\,{\sc ii} 3900 line can not be used for Al abundance determinations 
of B-type stars because of being too weak and heavily affected by blending.

\subsection{Trends of Al abundances in B-type stars}

In this subsection, the trends of Al abundances for normal and chemically peculiar 
stars are separately discussed. Here, we confine ourselves only to the non-LTE 
abundances derived from Al\,{\sc ii} 4663 line ($A^{\rm N}_{4663}$), which are
considered to be more reliable (Sect.~5.2.2).  

\subsubsection{Normal B-type stars}

The photospheric Al abundances of ordinary (non-CP) B-type stars should retain 
the composition of galactic gas from which they were formed.
As mentioned in Sect.~5.2.2, the non-LTE Al abundances ($A^{\rm N}_{4663}$) 
turn out to be almost independent upon $T_{\rm eff}$ over a wide range of B-type 
stars at 10000\,K\,$\la T_{\rm eff} \la 20000$\,K (Fig.~10d), accomplishing a reasonable
constancy. An inspection of Fig.~10d suggests that the demarcation line dividing
the non-CP group (homogeneous Al abundances) and CP group (considerably Al-deficient) 
may be set at $A = 6.0$. Then, those 36 stars (out of 51 stars) satisfying the criterion 
$A^{\rm N}_{4663} > 6.0$ are regarded as normal B stars, for which the mean abundance is 
calculated as $\langle A^{\rm N}_{4663} \rangle = 6.47$ (standard deviation is 
$\sigma =  0.15$). Alternatively, if two highest-$T_{\rm eff}$ stars (HD~000886 and 
HD~035708) showing some deviation from the main trend (cf. Sect.~5.2.2) are excluded, 
$\langle A^{\rm N}_{4663} \rangle = 6.45$ with $\sigma =  0.13$.

This $\langle A^{\rm N}_{4663} \rangle$ is in remarkable agreement with the solar 
abundance ($A_{\odot} = 6.47$). That is, Al abundances 
of young B-type stars (representing the gas composition at the time of 
some $\sim 10^{7}$--$10^{8}$\,yr ago) are almost similar to that of the Sun 
(formed $\sim 4.6 \times 10^{9}$\,yr ago), which may be regarded as reasonable.
Therefore, the trend of Al abundances in normal B-type stars is markedly different  
from the case of phosphorus (systematically subsolar by $\sim$\,0.2--0.3\,dex 
in contradiction with the standard concept of galactic chemical evolution; cf. Paper~II).  

\subsubsection{Peculiar HgMn stars}

According to the demarcation in Sect.~5.3.1, those 15 stars with $A^{\rm N}_{4663} < 6.0$ 
are Al-deficient CP stars (yellow-dotted symbols in Fig.~10), which are mostly classified 
as HgMn stars (non-magnetic late B-type chemically peculiar stars). These stars are 
considerably underabundant in Al relative to the Sun. The extent of deficiency is 
diversified from star to star (by $\sim$\,0.5--2.0\,dex cf. Fig.~10d), though a rough 
tendency is observed that the anomaly tends to become more conspicuous with an increase 
in $T_{\rm eff}$ (from [Al/H]\,$\sim -1$ around $T_{\rm eff} \sim 10000$\,K to 
[Al/H]\,$\sim -2$ around $T_{\rm eff} \sim 14000$\,K). 

This is almost a reconfirmation of the trend shown in Fig.~3 of Ghazaryan and Alecian 
(2016), though their figure (see ``Al\_teff.pdf'' included in their online material) 
indicates [Al/H] to be as low as $\sim -2.5$ (while the lowest [Al/H] in Fig.~10d is 
$\sim -2$), which might be due to the neglect of positive non-LTE corrections 
in previous determinations. 

\section{Summary and conclusion}

Our understanding of the photospheric abundances of aluminium in B-type stars 
is not sufficient, despite that they may provide us with important information 
about the composition of galactic gas at the time of their formation.
That is, previous abundance studies of this element seem to have been done   
more on chemically peculiar HgMn stars (late B-type), while those for relevant 
normal B-stars (retaining the original gas composition in their atmosphere) 
are comparatively scarce and do not seem to be sufficiently reliable 
because of the diversified results.

Motivated by this situation, a spectroscopic study was conducted to determine 
the aluminium abundances for 85 early-to-late B-type main-sequence stars 
(10000\,K\,$\la T_{\rm eff} \la$\,22000\,K) (comprising normal stars as well as 
chemically peculiar HgMn stars) were carried out by using two Al\,{\sc ii} 
lines at 6243 and 4663\,\AA. 

In order to take into account the non-LTE effect in the abundance analysis, 
non-LTE calculations were carried out on an extensive grid of models.
It turned out that the non-LTE effect for these Al\,{\sc ii} 6243/4663 lines  
acts in the direction of line-weakening (i.e., line profile becomes shallower) 
caused by an overionization-induced line opacity decrease along with an 
enhancement of line source function, and this effect becomes more conspicuous 
with an increase in $T_{\rm eff}$ as well as with a decrease in $\log g$.

As for the observational data of program stars used for abundance 
determination, the same orange-region spectra as used in Paper~II were adopted 
for the 6243 line, while the blue-region spectra available in the public-open 
database (ESO, CFHT, ELODIE) were employed for the 4663 line.  

The abundance determination was carried out by the two-step process: 
(1) A spectrum-fitting analysis was first applied to the wavelength regions 
of these two lines and their equivalent widths ($W_{6243}$ and $W_{4663}$) 
were derived from the fitting-based abundance solutions. (2) Then, non-LTE 
abundances/corrections  as well as possible errors were evaluated from 
such established $W$ values. The following conclusions are extracted from 
the results of Al abundances. 

Regarding the non-LTE abundances of normal stars resulting from the 
Al\,{\sc ii} 6243 line, some $T_{\rm eff}$-dependent systematic trend remains 
unremoved, which means that non-LTE corrections evaluated for this line are 
quantitatively insufficient, despite that our non-LTE calculation reasonably 
reproduces the qualitative behavior of this line (e.g., appearance of emission 
at the higher-$T_{\rm eff}$ regime of early B-type stars).

Meanwhile, for the case of the Al\,{\sc ii} 4663 line, which is more advantageous 
than the 6243 line because it is stronger without showing any emission,
the resulting non-LTE abundances of ordinary B stars are almost 
constant at the solar abundance ($A \simeq 6.5$) over the wide 
$T_{\rm eff}$ range ($\sim$\,10000--20000\,K), which suggests that the abundances 
derived from this line are successfully non-LTE-corrected and trustable.

Therefore, according to these results derived from the Al\,{\sc ii} 4663 line, we 
may reasonably state that the Al abundance of the galactic gas, from which 
early-to-late B-type stars were born several times $\sim 10^{7}$--$10^{8}$\,yr ago, 
is almost similar to the solar composition. This consequence is markedly different from 
the case of phosphorus (systematically subsolar by $\sim$\,0.2--0.3\,dex; cf. Paper~II).

As to the photospheric Al abundances of chemically peculiar HgMn stars 
(10000\,K\,$\la T_{\rm eff} \la 15000$\,K), our analysis resulted that this element is 
conspicuously underabundant by $\sim$\,0.5--2\,dex in comparison with the Sun (as well as 
normal B stars) and the extent of deficiency tends to increase towards higher $T_{\rm eff}$, 
which is a reconfirmation of the characteristics already reported in the past literature. 

\acknowledgements

This investigation has made use of the SIMBAD database, operated by CDS, 
Strasbourg, France, and the VALD database operated at Uppsala University,
the Institute of Astronomy RAS in Moscow, and the University of Vienna. 
This study is partly based on the data obtained from the ESO Science Archive Facility, 
ELODIE archive, and CFHT Science Archive Data (via CADC), as detailed in Table~3.

\newpage

\begin{figure}[H]
\centerline{\includegraphics[width=0.8\textwidth,clip=]{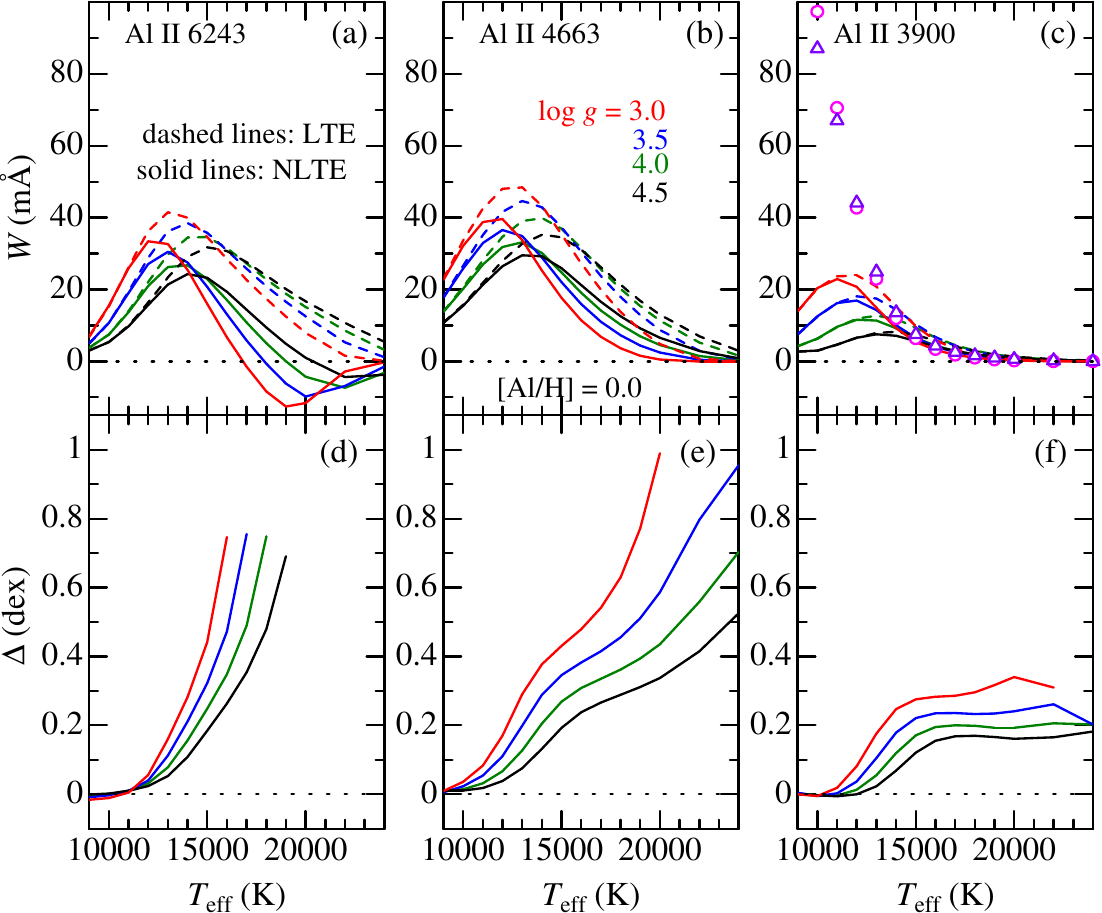}}
\caption{
Non-LTE and LTE equivalent widths ($W^{\rm N}$ and $W^{\rm L}$ depicted in
solid and dashed lines; upper panels (a)--(c)) and the corresponding 
non-LTE corrections ($\Delta$; lower panels (d)--(f)), which were 
computed with the solar Al abundance on the non-LTE 
grid of models described in Sect~2.1, are plotted against $T_{\rm eff}$.
Left ((a), (d)), center ((b), (e)), and right ((c), (f)) panels are 
for the Al\,{\sc ii} 6243, 4663, and 3900 lines, respectively. 
Results for different $\log g$ are distinguished by line colors 
(red, blue, green, black for 3.0, 3.5, 4.0, and 4.5). 
Note that $\Delta$ can not be calculated for the case of 
$W^{\rm N} \le 0$ (because LTE abundance is unable to determine). 
In panel~(c), the expected strengths of the Ti\,{\sc ii} 
3900.539 line (contaminating the Al\,{\sc ii} 3900 line), which were 
calculated with the solar Ti abundance ($A_{\odot}$(Ti) = 4.99) for 
$\log g = 3.0$ (pink open circles) and 4.0 (violet open triangles) models, 
are also plotted against $T_{\rm eff}$ for comparison.   
}
\label{fig1}
\end{figure}

\begin{figure}[H]
\centerline{\includegraphics[width=0.8\textwidth,clip=]{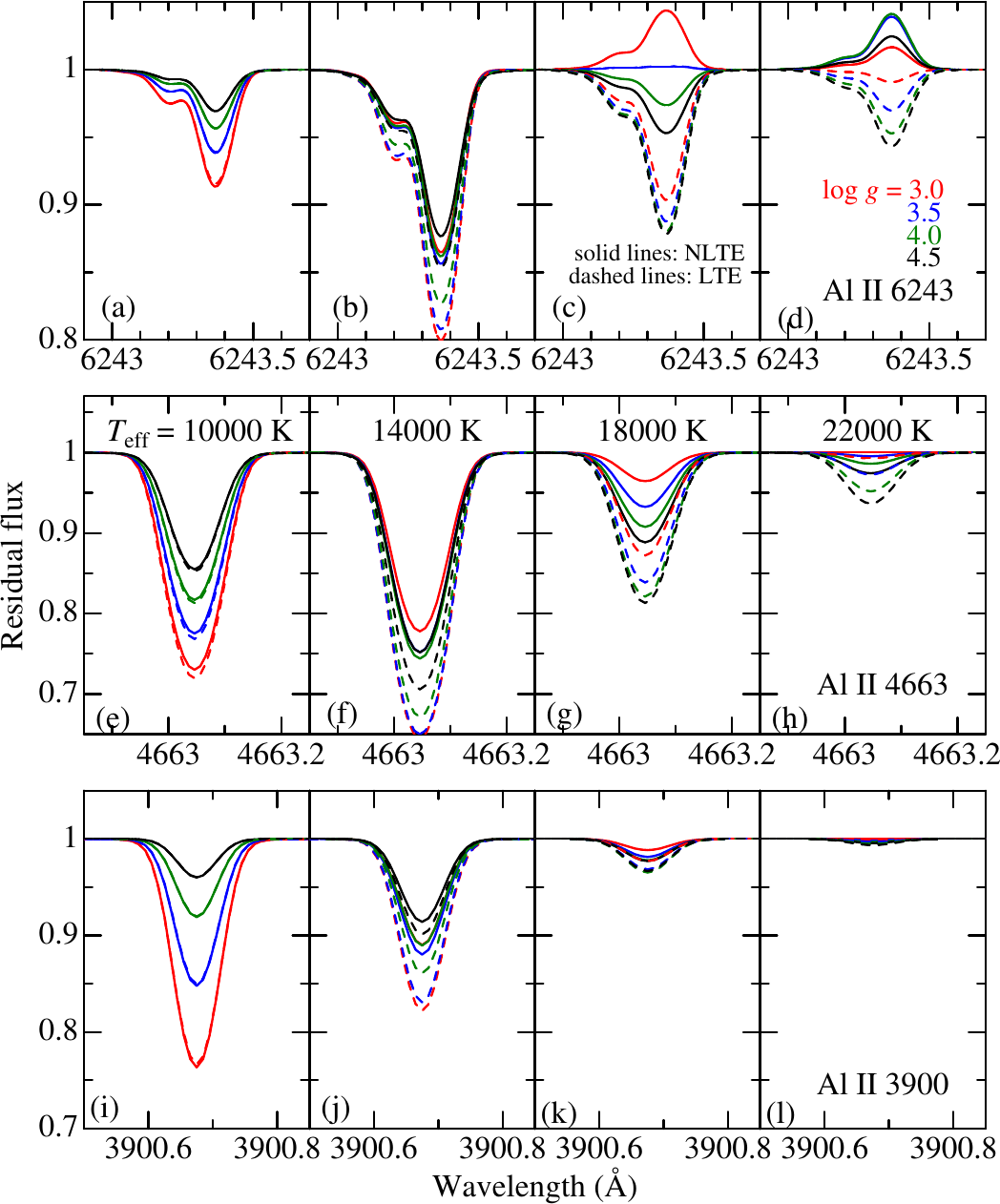}}
\caption{Theoretical line profiles computed for the Al\,{\sc ii} 6243 (top panels 
(a)--(d)), 4663 (middle panels (e)--(h)), and 3900 (bottom panels (i)--(l)) lines,
each corresponding to $T_{\rm eff}$ = 10000\,K, 14000\,K, 18000\,K, and 22000\,K
(from left to right). The same meanings of the line types and line colors
as in Fig.~1.  
}
\label{fig2}
\end{figure}

\begin{figure}[H]
\centerline{\includegraphics[width=0.8\textwidth,clip=]{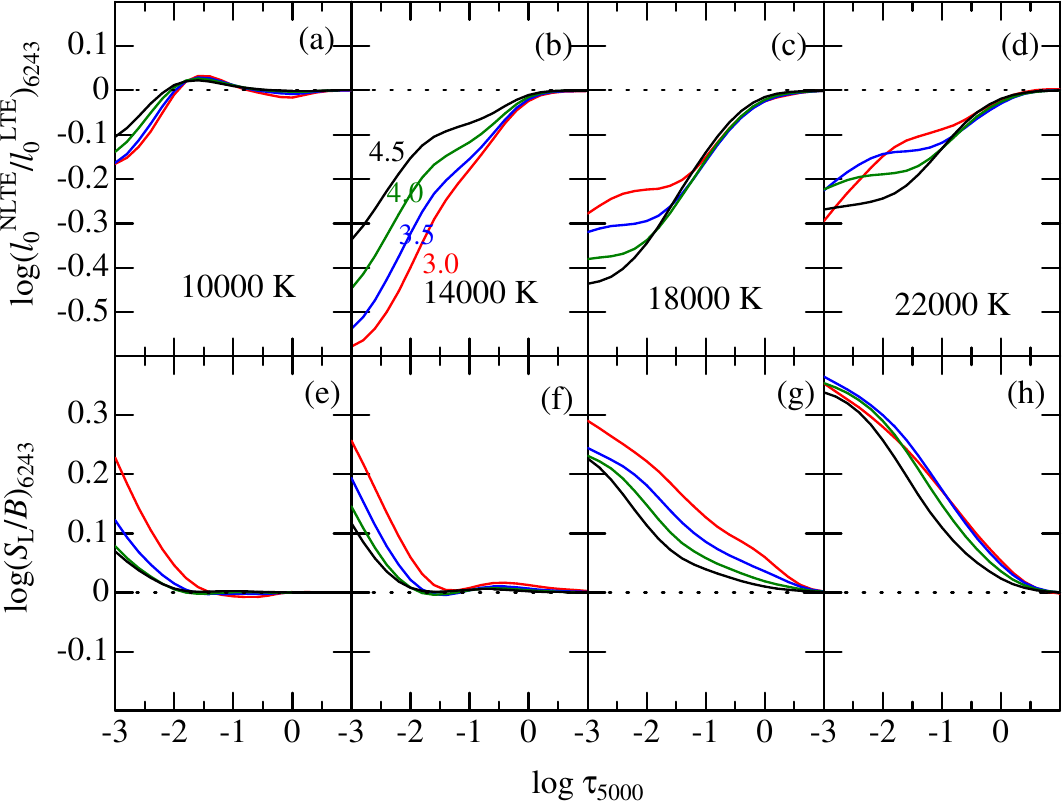}}
\caption{
The non-LTE-to-LTE line-center opacity ratio (upper panels (a)--(d)) and 
the ratio of the line source function ($S_{\rm L}$) 
to the local Planck function ($B$) (lower panels (e)--(f))  
for the Al\,{\sc ii} $^{3}$P$^{\rm o}$--$^{3}$D transition 
(corresponding to Al\,{\sc ii} 6243) of multiplet~10, 
plotted against the continuum optical depth at 5000\,\AA. 
Shown here are the calculations done with $v_{\rm t} = 2$\,km\,s$^{-1}$ 
and the solar Al abundance ([Al/H] = 0) for four representative $T_{\rm eff}$ 
values (from left to right): 10000\,K (panels (a), (e)), 14000\,K (panels (b), (f)), 
18000\,K (panels (c), (g)), and 22000\,K (panels (d), (h)).
The same meanings of the line colors as in Fig.~1.   
}
\label{fig3}
\end{figure}

\begin{figure}[H]
\centerline{\includegraphics[width=0.8\textwidth,clip=]{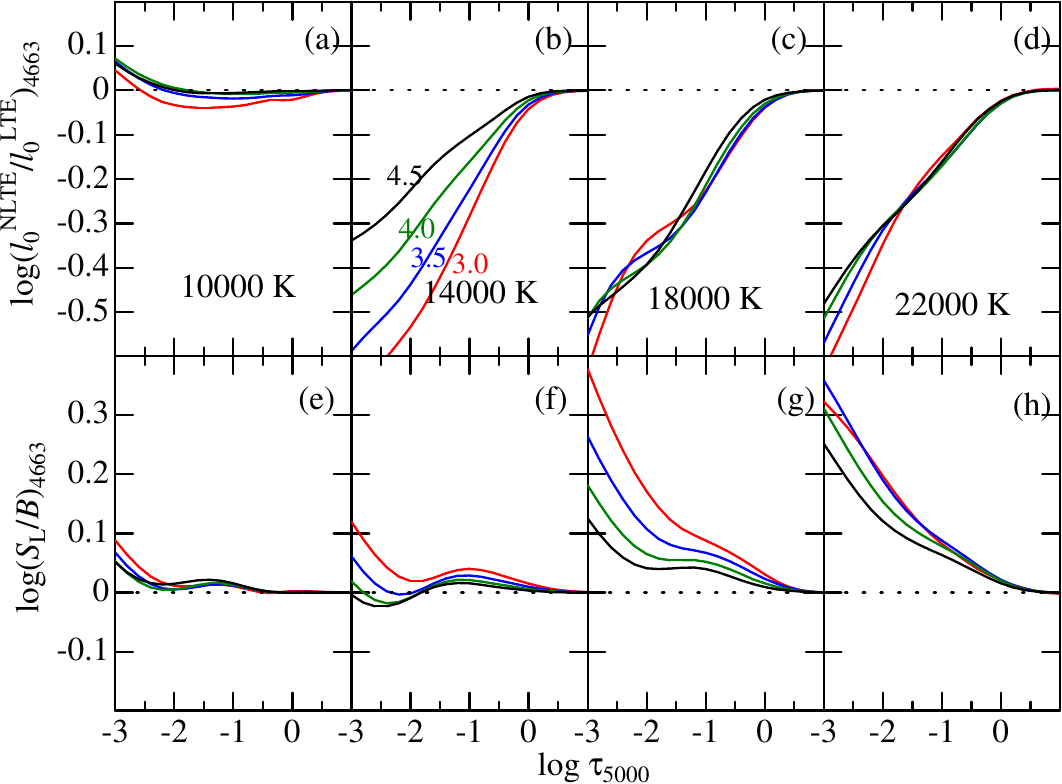}}
\caption{
The non-LTE-to-LTE line-center opacity ratio and 
the ratio of the line source function ($S_{\rm L}$) 
to the local Planck function ($B$)   
for the Al\,{\sc ii} $^{1}$D--$^{1}$P$^{\rm o}$ transition 
(corresponding to Al\,{\sc ii} 4663) of multiplet~2, 
plotted against the continuum optical depth at 5000\,\AA. 
Otherwise, the same as in Fig.~3.
}
\label{fig4}
\end{figure}

\begin{figure}[H]
\centerline{\includegraphics[width=0.8\textwidth,clip=]{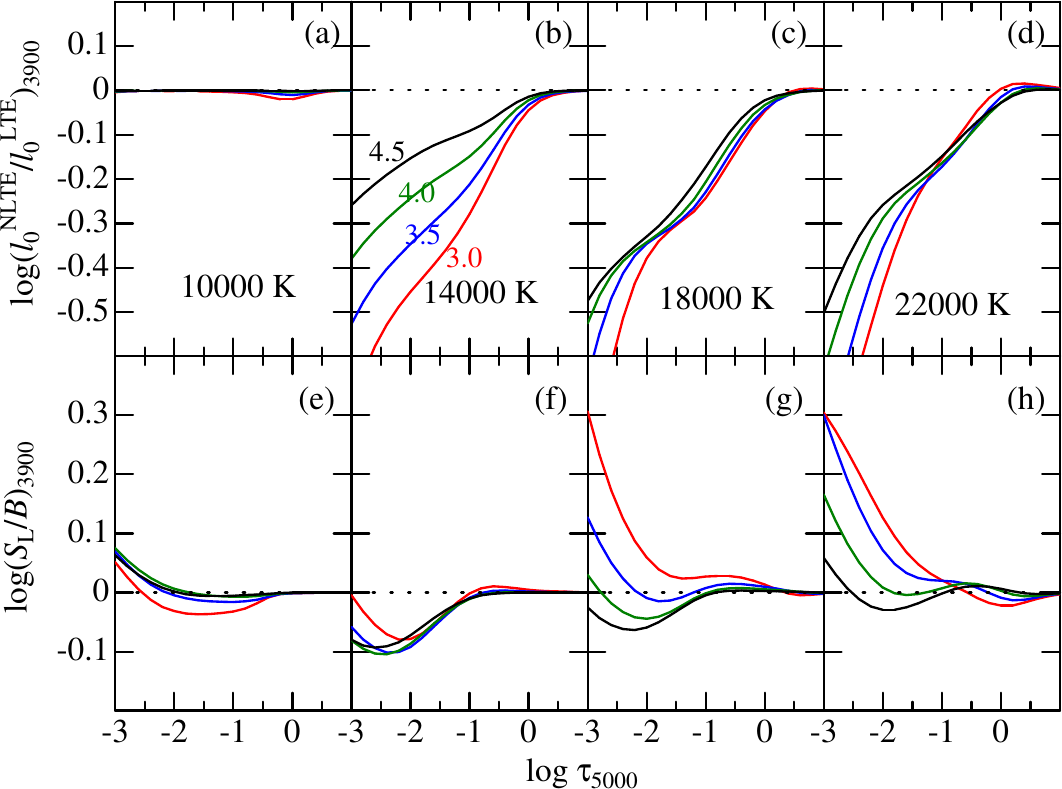}}
\caption{
The non-LTE-to-LTE line-center opacity ratio and 
the ratio of the line source function ($S_{\rm L}$) 
to the local Planck function ($B$)  
for the Al\,{\sc ii} $^{1}$P$^{\rm o}$--$^{1}$D transition 
(corresponding to Al\,{\sc ii} 3900) of multiplet~1, 
plotted against the continuum optical depth at 5000\,\AA. 
Otherwise, the same as in Fig.~3.
}
\label{fig5}
\end{figure}

\begin{figure}[H]
\centerline{\includegraphics[width=0.6\textwidth,clip=]{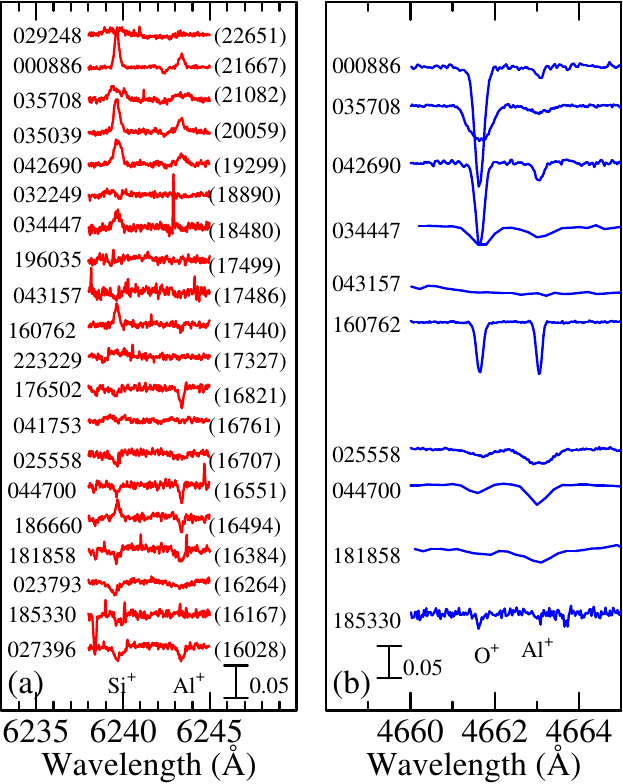}}
\caption{
Observed spectra of the early-to-mid B-type stars with 
$T_{\rm eff} \protect\ga 16000$\,K in the neighborhood of (a) Al\,{\sc ii} 
6243 line and (b) Al\,{\sc ii} 4663 line, 
which are designated by the HD number on the left and arranged in the 
descending order of $T_{\rm eff}$ (indicated by parenthesized values). 
Note that the Al\,{\sc ii} 6243 line (along with the neighboring Si\,{\sc ii} 
6239 line) shows an emission or a very weak strength due to filled-in 
emission at $T_{\rm eff} \protect\ga 17000$\,K, whereas the Al\,{\sc ii} 4663 line 
ever keeps an absorption profile.
}
\label{fig6}
\end{figure}

\begin{figure}[H]
\centerline{\includegraphics[width=0.9\textwidth,clip=]{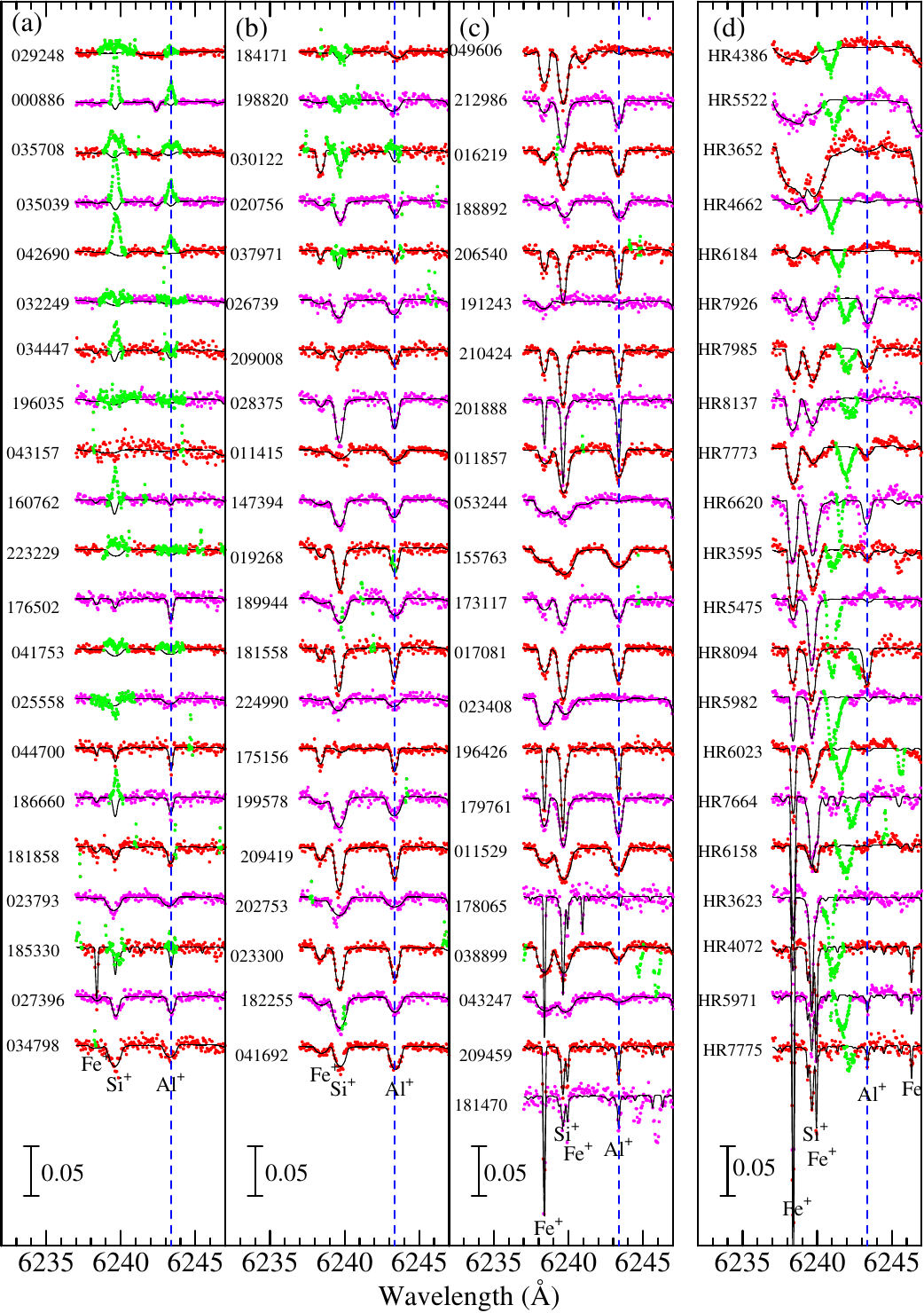}}
\caption{
Synthetic spectrum-fitting in the neighborhood of the Al\,{\sc ii} 6243 line 
done for each of the 85 program stars. 
Shown here is the selected wavelength region of 6237--6247\,\AA.
The best-fit theoretical spectra are depicted by black solid lines, 
while the observed data are plotted by colored dots (outlier data points 
rejected in calculating $\chi^{2}$ are highlighted in light green).  
The spectra are arranged in the same manner as Fig.~3 of Paper~II
to keep consistency with that paper: Panels (a), (b), and (c) are for 
the 64 early-to-late B stars (indicated by the HD number) in the descending order 
of $T_{\rm eff}$ (from top to bottom; from left to right). while the rightmost 
panel (d) is for 21 late-B stars (indicated by the HR number) in the descending 
order of $v_{\rm e}\sin i$. An offset of 0.05 (in unit of the continuum) 
is applied to each spectrum relative to the adjacent one. 
The position of 6243.367\,\AA\ (wavelength of the strongest component of
Al\,{\sc ii} 6243 triplet) is shown by the vertical dashed line.
The wavelength scale is in the laboratory frame after correcting 
the radial velocity shift.  
}
\label{fig7}
\end{figure}

\begin{figure}[H]
\centerline{\includegraphics[width=0.9\textwidth,clip=]{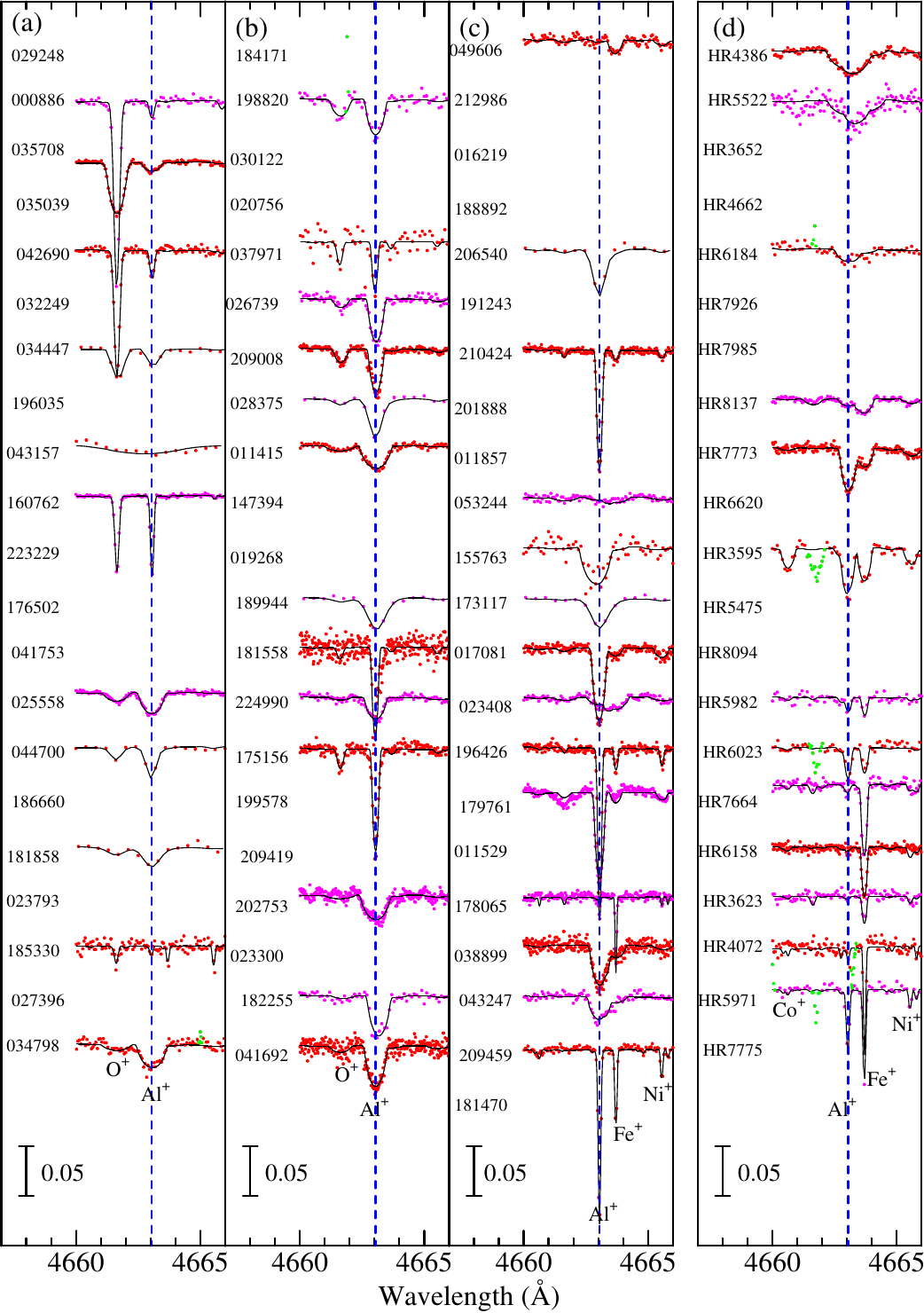}}
\caption{
Synthetic spectrum-fitting in the neighborhood of the Al\,{\sc ii} 4663 line
done for 51 stars, for which public-open data are available (cf. Table~3). 
Shown here is the selected wavelength region of 4660--4666\,\AA.
The position of 4663.046\,\AA\ (wavelength of the 
Al\,{\sc ii} 4663 line) is shown by the vertical dashed line.
Otherwise, the same as in Fig.~7.
}
\label{fig8}
\end{figure}

\begin{figure}[H]
\centerline{\includegraphics[width=0.7\textwidth,clip=]{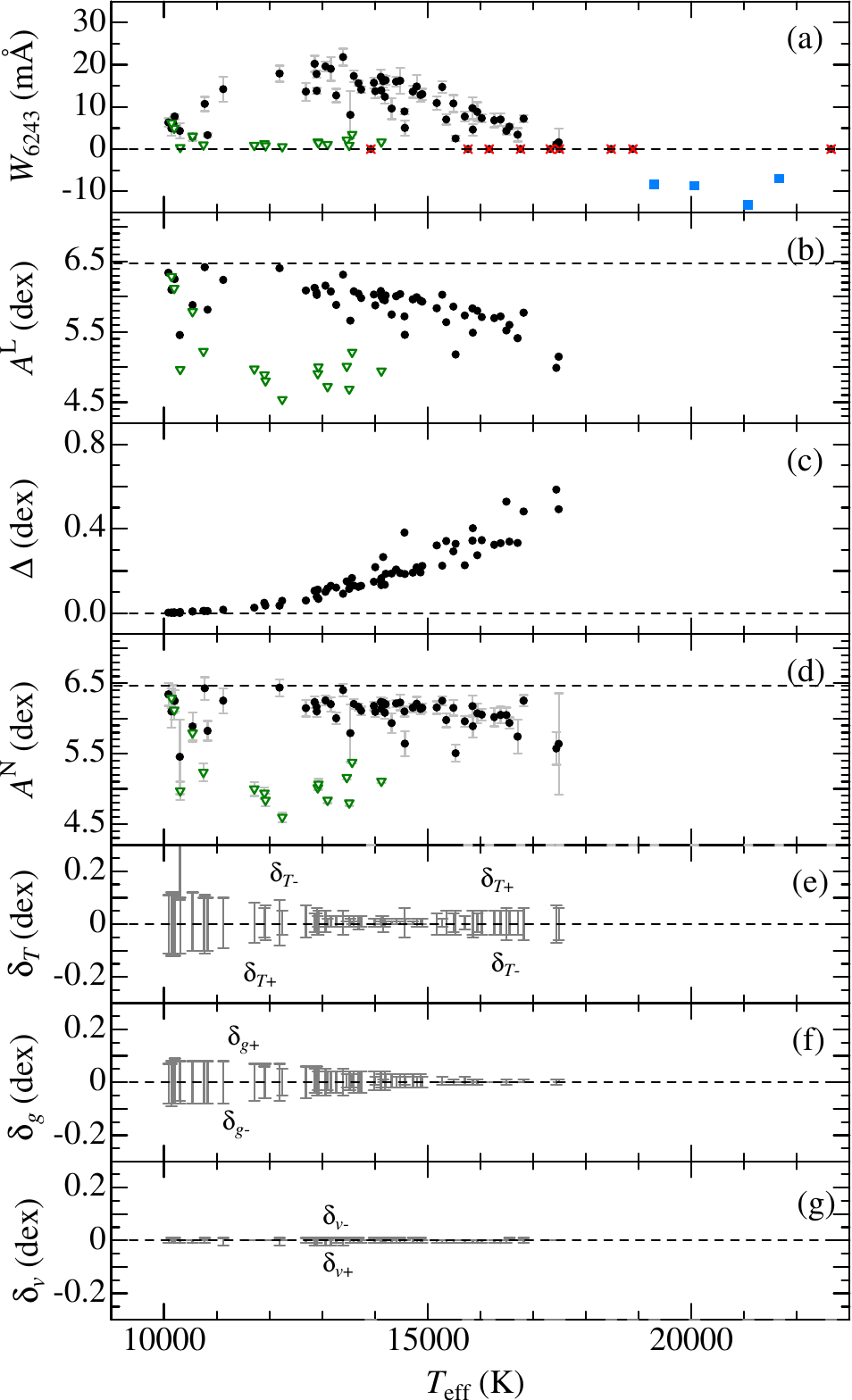}}
\caption{
Abundance-related quantities for the Al\,{\sc ii}~6243 line are plotted against $T_{\rm eff}$.  
(a) $W_{6243}$ (equivalent width), where the indicated error bars ($\pm\delta W$) 
are their uncertainties (see Sect.~4.4).
(b) $A^{\rm L}$ (LTE aluminium abundance).
(c) $\Delta$ ($\equiv A^{\rm N} - A^{\rm L}$; non-LTE correction).
(d) $A^{\rm N}$ (non-LTE aluminium abundance)
Here, the attached error bars are the root-sum-squares of $\delta_{W}$
(abundance ambiguities corresponding to $\delta W$), 
$\delta_{T}$, $\delta_{g}$, and $\delta_{v}$.
(e) $\delta_{T+}$ and $\delta_{T-}$ (abundance variations 
in response to $T_{\rm eff}$ changes of $+3 \%$ and $-3 \%$). 
(f) $\delta_{g+}$ and $\delta_{g-}$ (abundance variations 
in response to $\log g$ changes of $+0.2$\,dex and $-0.2$\,dex). 
(g) $\delta_{v+}$ and $\delta_{v-}$ (abundance variations 
in response to perturbing the standard $v_{\rm t}$ value by $\pm 1$\,km\,s$^{-1}$). 
Note that the signs of $\delta_{T}$ and $\delta_{g}$ are reversed on 
both sides of $T_{\rm eff}$ around $\sim$\,15000\,K.
In panels (a), (b), and (d), the three cases where $W$ or $A$ values could not orderly 
be established (cf. Sect.~4.3 and 4.4) are distinguished by differently colored symbols: 
blue squares $\cdots$ (negative) $W$ values directly measured by Gaussian fitting
(appreciable emission-line case (i));
overplotted red crosses $\cdots$ tentatively assigned zero values (too weak line 
case (ii) due to filled-in emission in mid- to early-B stars); 
green open downward triangle $\cdots$ upper limit values (too weak line case (iii) 
due to Al deficiency in late B-type HgMn stars). 
}
\label{fig9}
\end{figure}

\begin{figure}[H]
\centerline{\includegraphics[width=0.7\textwidth,clip=]{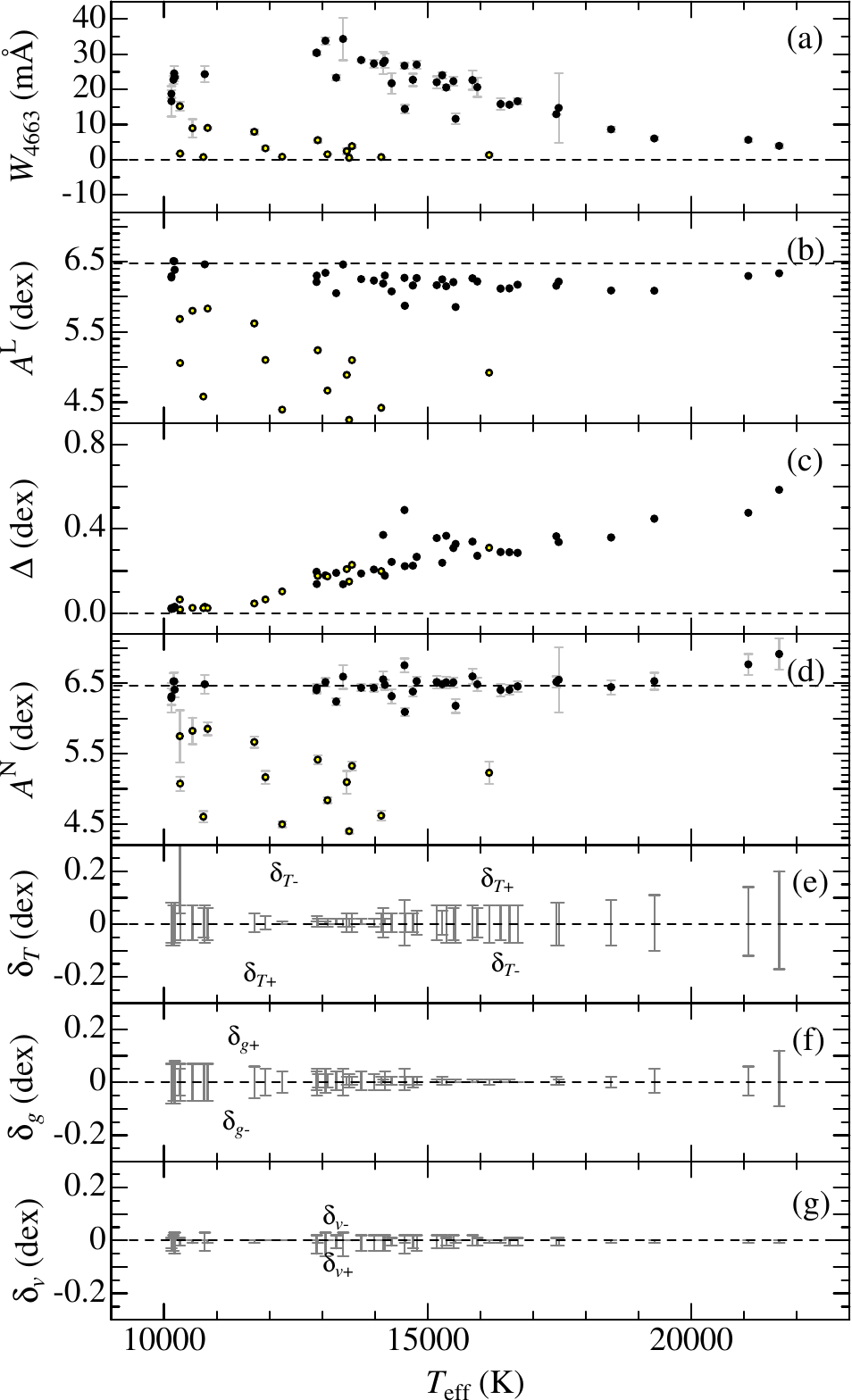}}
\caption{
Abundance-related quantities for the Al\,{\sc ii}~4663 line (equivalent widths, 
LTE/NLTE abundances as well as NLTE corrections, and sensitivities due to 
perturbations of atmospheric parameters) are plotted against $T_{\rm eff}$. 
In panels (a)--(d), 15 chemically peculiar stars with $A^{\rm N}_{4663} < 6.0$ 
are distinguished by overplotting \textcolor{red}{yellow small filled circles} 
on the symbols. 
Otherwise, the same as in Fig.~9.
}
\label{fig10}
\end{figure}

\begin{figure}[H] 
\centerline{\includegraphics[width=0.8\textwidth,clip=]{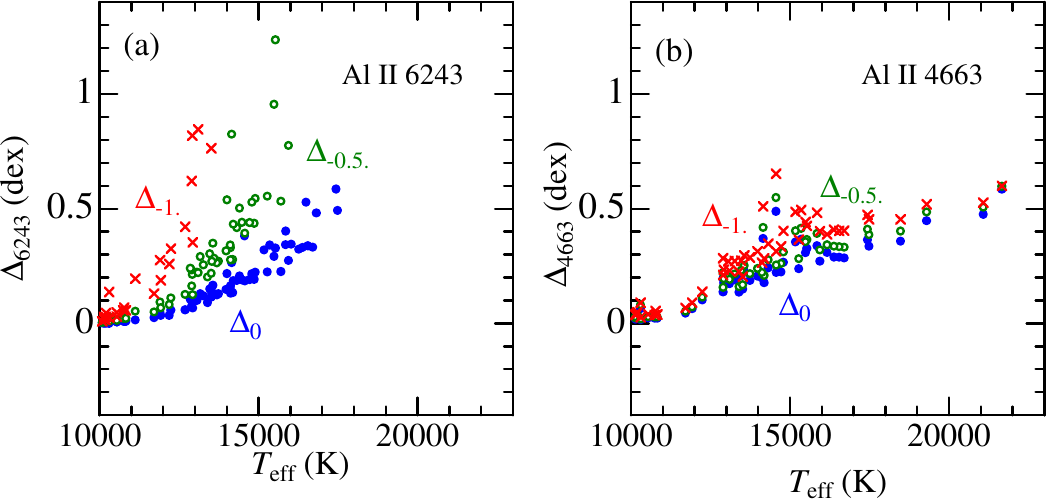}}
\caption{
Test of how the non-LTE abundance correction for each star (derived from a given equivalent width)
depends upon the Al abundance assumed in the statistical equilibrium calculations. 
Three kinds of non-LTE corrections ($\Delta_{0}$, $\Delta_{-0.5}$, $\Delta_{-1}$; 
depicted by filled symbols, open symbols, and crosses, respectively) are plotted 
against $T_{\rm eff}$, each of which correspond to different Al abundances adopted 
in non-LTE calculations ([Al/H] = 0, $-0.5$, and $-1.0$). Panels (a) and (b) are for 
Al\,{\sc ii} 6243 and Al\,{\sc ii} 4663, respectively. 
}
\label{fig11}
\end{figure}

\begin{figure}[H]
\centerline{\includegraphics[width=0.8\textwidth,clip=]{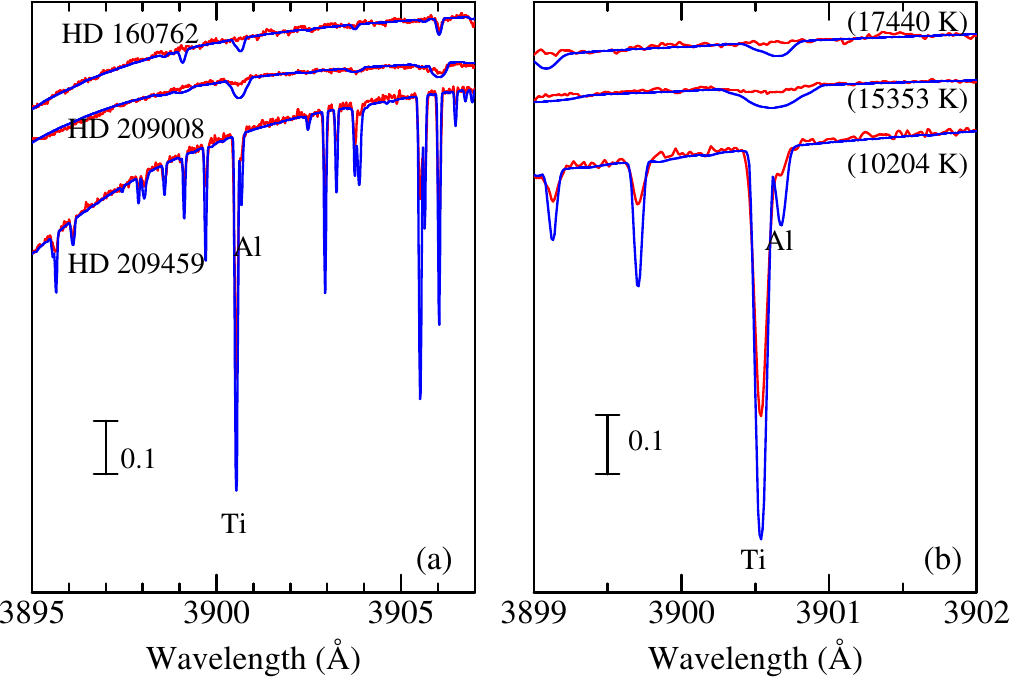}}
\caption{
Observed spectra of HD~160762, HD~209008, and HD~209459 around the $\sim 3900$\,\AA\ 
region comprising the Ti\,{\sc ii} 3900.539 and Al\,{\sc ii} 3900.675 lines
(red lines) are compared with the theoretical spectra 
(blue lines) calculated with the solar Ti and Al abundances ($A_{\odot}$(Ti) = 4.99 
and $A_{\odot}$(Al) = 6.47) under the assumption of LTE. 
Note that this region is on the red wing of H$\zeta$ 3889. 
Panels (a) and (b) are the wide view (3895--3907\,\AA) and the narrow view 
(3899--3902\,\AA), respectively.
}
\label{fig12}
\end{figure}

\newpage

\setcounter{table}{0}
\begin{table}[H]
\scriptsize
\caption{Atomic data of important lines.}
\begin{center}
\begin{tabular}
{ccccccccc}\hline \hline
Species & Multiplet & Transition & $\lambda$ & $\chi_{\rm low}$ & $\log gf$ & Gammar & Gammas & Gammaw \\
         &  No.     &       & (\AA)  & (eV)  &  (dex)  & (dex)  & (dex)  &  (dex) \\
\hline
Ti\,{\sc ii} & 34 &  a$^{2}$G$_{9/2}$--z$^{2}$G$^{\rm o}_{9/2}$ & 3900.539 & 1.131 & $-0.290$ & 8.31 & $-6.49$ & $-7.82$   \\
Al\,{\sc ii} & 1  &  $^{1}$P$^{\rm o}_{1}$--$^{1}$D$_{2}$       & 3900.675 & 7.421 & $-2.26^{*}$ & 9.22 & $(-5.95)$ & $(-7.77)$ \\
\hline
Al\,{\sc ii} & 2  & $^{1}$D$_{2}$--$^{1}$P$^{\rm o}_{1}$        & 4663.046 &10.598 & $-0.284$ & 7.99 & $(-5.53)$ & $(-7.64)$ \\
\hline
Si\,{\sc ii} & -- & $^{2}$F$^{\rm o}_{7/2}$--$^{2}$G$_{7/2}$    & 6239.614 &12.840 & $-1.359$ & 8.59 & $-3.54$   & $-7.25$   \\
             &    & $^{2}$F$^{\rm o}_{7/2}$--$^{2}$G$_{9/2}$    & 6239.614 &       & $+0.185$ &      &           &           \\
             &    & $^{2}$F$^{\rm o}_{5/2}$--$^{2}$G$_{7/2}$    & 6239.665 &       & $+0.072$ &      &           &           \\
Al\,{\sc ii} & 10 & $^{3}$P$^{\rm o}_{2}$--$^{3}$D$_{1}$        & 6243.073 &13.077 & $-1.250$ &(7.76)& $-4.76$   & $(-7.50)$ \\
             &    & $^{3}$P$^{\rm o}_{2}$--$^{3}$D$_{2}$        & 6243.203 &       & $-0.080$ &      &           &           \\
             &    & $^{3}$P$^{\rm o}_{2}$--$^{3}$D$_{3}$        & 6243.367 &       & $+0.670$ &      &           &           \\
\hline
\end{tabular}
\end{center}
\scriptsize
All these data are taken from the VALD database (Ryabchikova {\it et al.}, 2015),
except for the $\log gf$ value of the Al\,{\sc ii} 3900.675 line, for which the 
NIST data was adopted instead (because the VALD value is likely to be unreliable; 
cf. Sect.~5.2.3).
Note that the Al\,{\sc ii} 6243 line comprises three components belonging to 
the same multiplet.  The data of the Ti\,{\sc ii} 3900.539 line (contaminating 
Al\,{\sc ii} 3900) and the Si\,{\sc ii} 6239 line (high-excitation line 
similar to Al\,{\sc ii} 6243) are also shown.
After the first six self-explanatory columns, damping parameters are given 
in the last three columns, where the parenthesized values are calculated
by Kurucz's (1993) WIDTH9 program (by following the default treatment)
because the data are not given in VALD:\\
Gammar is the radiation damping width (s$^{-1}$), $\log\gamma_{\rm rad}$.\\
Gammas is the Stark damping width (s$^{-1}$) per electron density (cm$^{-3}$) 
at $10^{4}$ K, $\log(\gamma_{\rm e}/N_{\rm e})$.\\
Gammaw is the van der Waals damping width (s$^{-1}$) per hydrogen density 
(cm$^{-3}$) at $10^{4}$ K, $\log(\gamma_{\rm w}/N_{\rm H})$. \\
$^{*}$Taken from the NIST database.  
\end{table}

\setcounter{table}{1}
\begin{table}[H]
\scriptsize
\caption{Program stars and the results of analysis.}
\begin{center}
\begin{tabular}{cccccc r@{ }r@{ }r r@{ }r@{ }r l}\hline\hline
No. & HD\# & HR\# & $T_{\rm eff}$ & $\log g$  & $v_{\rm e}\sin i$ & 
$W_{6243}$ & $A^{\rm N}_{6243}$ & $\Delta_{6243}$ & 
$W_{4663}$ & $A^{\rm N}_{4663}$ & $\Delta_{4663}$ & Remark \\
(1) & (2) & (3) & (4) & (5) & (6) & (7) & (8) & (9) & (10) & (11) & (12) & (13) \\
\hline
 1 & 029248 & 1463 & 22651 & 3.58 &  46 &    [0.0]&$\cdots$&$\cdots$& \multicolumn{3}{c}{(no obs. data)} & (ii) \\
 2 & 000886 & 0039 & 21667 & 3.83 &   9 &  $-7.0$ &$\cdots$&$\cdots$&   3.9 &  6.92 &  0.58 & (i) \\
 3 & 035708 & 1810 & 21082 & 4.09 &  26 & $-13.2$ &$\cdots$&$\cdots$&   5.6 &  6.77 &  0.47 & (i) \\
 4 & 035039 & 1765 & 20059 & 3.69 &  10 &  $-8.7$ &$\cdots$&$\cdots$& \multicolumn{3}{c}{(no obs. data)} & (i) \\
 5 & 042690 & 2205 & 19299 & 3.81 &  12 &  $-8.3$ &$\cdots$&$\cdots$&   6.0 &  6.53 &  0.45 & (i) \\
 6 & 032249 & 1617 & 18890 & 4.13 &  40 &    [0.0]&$\cdots$&$\cdots$& \multicolumn{3}{c}{(no obs. data)} & (ii) \\
 7 & 034447 & 1731 & 18480 & 4.10 &   9 &    [0.0]&$\cdots$&$\cdots$&   8.6 &  6.44 &  0.36 & (ii) \\
 8 & 196035 & 7862 & 17499 & 4.36 &  35 &    [0.0]&$\cdots$&$\cdots$& \multicolumn{3}{c}{(no obs. data)} & (ii) \\
 9 & 043157 & 2224 & 17486 & 4.12 &  37 &     1.6 &  5.64  &  0.49  &  14.7 &  6.55 &  0.34 & \\
10 & 160762 & 6588 & 17440 & 3.91 &   7 &     1.1 &  5.57  &  0.58  &  12.9 &  6.52 &  0.36 & \\
11 & 223229 & 9011 & 17327 & 4.20 &  31 &    [0.0]&$\cdots$&$\cdots$& \multicolumn{3}{c}{(no obs. data)} & (ii) \\
12 & 176502 & 7179 & 16821 & 3.89 &   8 &     7.2 &  6.25  &  0.48  & \multicolumn{3}{c}{(no obs. data)} & \\
13 & 041753 & 2159 & 16761 & 3.90 &  28 &    [0.0]&$\cdots$&$\cdots$& \multicolumn{3}{c}{(no obs. data)} & (ii) \\
14 & 025558 & 1253 & 16707 & 4.29 &  41 &     3.4 &  5.74  &  0.33  &  16.6 &  6.45 &  0.28 & \\
15 & 044700 & 2292 & 16551 & 4.21 &   5 &     5.3 &  5.94  &  0.34  &  15.6 &  6.41 &  0.29 & \\
16 & 186660 & 7516 & 16494 & 3.57 &   9 &     4.3 &  6.05  &  0.53  & \multicolumn{3}{c}{(no obs. data)} & \\
17 & 181858 & 7347 & 16384 & 4.19 &  17 &     7.0 &  6.05  &  0.33  &  15.8 &  6.40 &  0.29 & \\
18 & 023793 & 1174 & 16264 & 4.15 &  46 &     6.8 &  6.02  &  0.32  & \multicolumn{3}{c}{(no obs. data)} & \\
19 & 185330 & 7467 & 16167 & 3.77 &   4 &    [0.0]&$\cdots$&$\cdots$&   1.3 &  5.23 &  0.31 & (ii) \\
20 & 027396 & 1350 & 16028 & 3.91 &  15 &     7.3 &  6.05  &  0.34  & \multicolumn{3}{c}{(no obs. data)} & \\
21 & 034798 & 1753 & 15943 & 4.27 &  37 &     8.8 &  6.07  &  0.27  &  20.6 &  6.49 &  0.27 & \\
22 & 184171 & 7426 & 15858 & 3.54 &  27 &     4.6 &  5.89  &  0.40  & \multicolumn{3}{c}{(no obs. data)} & \\
23 & 198820 & 7996 & 15852 & 3.86 &  32 &     9.7 &  6.18  &  0.34  &  22.6 &  6.60 &  0.34 & \\
24 & 030122 & 1512 & 15765 & 3.72 &  15 &    [0.0]&$\cdots$&$\cdots$& \multicolumn{3}{c}{(no obs. data)} & (ii) \\
25 & 020756 & 1005 & 15705 & 4.43 &  18 &     7.7 &  5.96  &  0.23  & \multicolumn{3}{c}{(no obs. data)} & \\
26 & 037971 & 1962 & 15532 & 3.63 &   9 &     2.5 &  5.51  &  0.33  &  11.6 &  6.18 &  0.33 & \\
27 & 026739 & 1312 & 15490 & 3.92 &  31 &    10.8 &  6.15  &  0.29  &  22.3 &  6.51 &  0.31 & \\
28 & 209008 & 8385 & 15353 & 3.50 &  20 &     7.0 &  5.98  &  0.34  &  20.5 &  6.51 &  0.37 & \\
29 & 028375 & 1415 & 15278 & 4.30 &  19 &    14.7 &  6.25  &  0.22  &  24.0 &  6.49 &  0.24 & \\
30 & 011415 & 0542 & 15174 & 3.54 &  42 &    10.9 &  6.16  &  0.32  &  22.0 &  6.52 &  0.35 & \\
31 & 147394 & 6092 & 14898 & 4.01 &  30 &    13.0 &  6.15  &  0.22  & \multicolumn{3}{c}{(no obs. data)} & \\
32 & 019268 & 0930 & 14866 & 4.24 &  17 &    12.8 &  6.13  &  0.19  & \multicolumn{3}{c}{(no obs. data)} & \\
33 & 189944 & 7656 & 14793 & 4.01 &  35 &    14.8 &  6.21  &  0.22  &  27.0 &  6.53 &  0.27 & \\
34 & 181558 & 7339 & 14721 & 4.15 &  14 &    13.6 &  6.15  &  0.19  &  22.7 &  6.38 &  0.22 & \\
35 & 224990 & 9091 & 14569 & 3.99 &  35 &     5.0 &  5.64  &  0.19  &  14.4 &  6.09 &  0.22 & \\
36 & 175156 & 7119 & 14561 & 2.79 &  12 &     8.9 &  6.10  &  0.38  &  26.7 &  6.76 &  0.49 & \\
37 & 199578 & 8022 & 14480 & 4.02 &  27 &    16.2 &  6.23  &  0.19  & \multicolumn{3}{c}{(no obs. data)} & \\
38 & 209419 & 8403 & 14404 & 3.82 &  16 &    16.0 &  6.21  &  0.21  & \multicolumn{3}{c}{(no obs. data)} & \\
39 & 202753 & 8141 & 14318 & 3.84 &  40 &     9.6 &  5.93  &  0.19  &  21.7 &  6.32 &  0.24 & \\
40 & 023300 & 1141 & 14207 & 3.84 &  19 &    16.3 &  6.20  &  0.19  & \multicolumn{3}{c}{(no obs. data)} & \\
41 & 182255 & 7358 & 14190 & 4.29 &  28 &    12.4 &  6.08  &  0.13  &  28.1 &  6.48 &  0.18 & \\
42 & 041692 & 2154 & 14157 & 3.19 &  28 &    16.1 &  6.22  &  0.26  &  27.5 &  6.55 &  0.37 & \\
43 & 049606 & 2519 & 14121 & 3.82 &  19 &    ($<1.7$)& ($<5.1$) &  0.16  &   0.7 &  4.62 &  0.20 & (iii) \\
44 & 212986 & 8554 & 14121 & 4.27 &  20 &    13.9 &  6.14  &  0.13  & \multicolumn{3}{c}{(no obs. data)} & \\
45 & 016219 & 0760 & 14113 & 4.06 &  23 &    17.1 &  6.23  &  0.15  & \multicolumn{3}{c}{(no obs. data)} & \\
46 & 188892 & 7613 & 14008 & 3.38 &  30 &    13.7 &  6.09  &  0.22  & \multicolumn{3}{c}{(no obs. data)} & \\
47 & 206540 & 8292 & 13981 & 4.01 &  13 &    15.7 &  6.18  &  0.15  &  27.3 &  6.43 &  0.21 & \\
48 & 191243 & 7699 & 13923 & 2.50 &  28 &    [0.0]&$\cdots$&$\cdots$& \multicolumn{3}{c}{(no obs. data)} & (ii) \\
49 & 210424 & 8452 & 13740 & 3.99 &  12 &    14.1 &  6.11  &  0.13  &  28.3 &  6.44 &  0.19 & \\
50 & 201888 & 8109 & 13689 & 4.01 &   5 &    15.6 &  6.16  &  0.12  & \multicolumn{3}{c}{(no obs. data)} & \\
51 & 011857 & 0561 & 13600 & 3.88 &  20 &    17.3 &  6.21  &  0.13  & \multicolumn{3}{c}{(no obs. data)} & \\
52 & 053244 & 2657 & 13467 & 3.42 &  36 &    ($<2.2$)& ($<5.2)$  &  0.15  &   2.4 &  5.09 &  0.21 & (iii) \\
53 & 155763 & 6396 & 13397 & 4.24 &  41 &    21.8 &  6.40  &  0.09  &  34.3 &  6.59 &  0.14 & \\
54 & 173117 & 7035 & 13267 & 3.63 &  22 &    12.7 &  6.00  &  0.12  &  23.3 &  6.24 &  0.19 & \\
55 & 017081 & 0811 & 13063 & 3.72 &  20 &    19.6 &  6.26  &  0.10  &  33.8 &  6.52 &  0.18 & \\
56 & 023408 & 1149 & 12917 & 3.36 &  30 &    ($<1.7$)& ($<5.0$) &  0.11  &   5.5 &  5.41 &  0.18 & (iii) \\
57 & 196426 & 7878 & 12899 & 3.89 &   6 &    13.8 &  6.10  &  0.08  &  30.3 &  6.43 &  0.14 & \\
58 & 179761 & 7287 & 12895 & 3.46 &  16 &    17.8 &  6.16  &  0.11  &  30.4 &  6.40 &  0.19 & \\
59 & 011529 & 0548 & 12858 & 3.43 &  30 &    20.2 &  6.23  &  0.10  & \multicolumn{3}{c}{(no obs. data)} & \\
60 & 178065 & 7245 & 12243 & 3.49 &   4 &    ($<0.6$)& ($<4.6$) &  0.06  &   0.8 &  4.49 &  0.10 & (iii) \\
61 & 038899 & 2010 & 10774 & 4.02 &  26 &    10.7 &  6.43  &  0.01  &  24.3 &  6.49 &  0.03 & \\
62 & 043247 & 2229 & 10301 & 2.39 &  33 &     4.3 &  5.45  &  0.00  &  15.2 &  5.75 &  0.06 & \\
63 & 209459 & 8404 & 10204 & 3.53 &   3 &     7.7 &  6.25  &  0.00  &  23.4 &  6.41 &  0.03 & \\
64 & 181470 & 7338 & 10085 & 3.92 &   2 &     6.3 &  6.34  &  0.00  & \multicolumn{3}{c}{(no obs. data)} & \\
\hline
\end{tabular} 
\end{center}
\end{table}

\setcounter{table}{1}
\begin{table}[H]
\scriptsize
\caption{(Continued.)}
\begin{center}
\begin{tabular}{cccccc r@{ }r@{ }r r@{ }r@{ }r l}\hline\hline
No. & HD\# & HR\# & $T_{\rm eff}$ & $\log g$  & $v_{\rm e}\sin i$ & 
$W_{6243}$ & $A^{\rm N}_{6243}$ & $\Delta_{6243}$ & 
$W_{4663}$ & $A^{\rm N}_{4663}$ & $\Delta_{4663}$ & Remark \\
(1) & (2) & (3) & (4) & (5) & (6) & (7) & (8) & (9) & (10) & (11) & (12) & (13) \\
\hline
65 & 098664 & 4386 & 10194 & 3.75 &  62 &    ($<5.0$)& ($<6.1$) &  0.00  &  24.5 &  6.53 &  0.03 & (iii) \\
66 & 130557 & 5522 & 10142 & 3.85 &  55 &    ($<6.2$)& ($<6.3$) &  0.00  &  16.6 &  6.29 &  0.02 & (iii) \\
67 & 079158 & 3652 & 13535 & 3.72 &  46 &     8.1 &  5.79  &  0.13  & \multicolumn{3}{c}{(no obs. data)} & \\
68 & 106625 & 4662 & 11902 & 3.36 &  37 &    ($<1.3$)& ($<4.9$) &  0.05  & \multicolumn{3}{c}{(no obs. data)} & (iii) \\
69 & 150100 & 6184 & 10542 & 3.84 &  36 &    ($<3.1$)& ($<5.8$) &  0.01  &   8.9 &  5.82 &  0.02 & (iii) \\
70 & 197392 & 7926 & 13166 & 3.46 &  30 &    19.0 &  6.20  &  0.13  & \multicolumn{3}{c}{(no obs. data)} & \\
71 & 198667 & 7985 & 11125 & 3.42 &  26 &    14.2 &  6.25  &  0.01  & \multicolumn{3}{c}{(no obs. data)} & \\
72 & 202671 & 8137 & 13566 & 3.36 &  25 &    ($<3.5$)& ($<5.4$) &  0.17  &   3.8 &  5.32 &  0.23 & (iii) \\
73 & 193432 & 7773 & 10180 & 3.91 &  24 &     5.9 &  6.27  &  0.00  &  22.7 &  6.53 &  0.02 & \\
74 & 161701 & 6620 & 12692 & 4.04 &  20 &    13.6 &  6.15  &  0.06  & \multicolumn{3}{c}{(no obs. data)} & \\
75 & 077350 & 3595 & 10141 & 3.68 &  20 &     4.9 &  6.10  &  0.00  &  18.7 &  6.31 &  0.02 & \\
76 & 129174 & 5475 & 12929 & 4.02 &  16 &    ($<1.5$)& ($<5.1$) &  0.07  & \multicolumn{3}{c}{(no obs. data)} & (iii) \\
77 & 201433 & 8094 & 12193 & 4.24 &  15 &    17.9 &  6.44  &  0.04  & \multicolumn{3}{c}{(no obs. data)} & \\
78 & 144206 & 5982 & 11925 & 3.79 &  12 &    ($<0.8$)& ($<4.8$) &  0.04  &   3.2 &  5.16 &  0.06 & (iii) \\
79 & 145389 & 6023 & 11714 & 4.02 &  11 &    ($<0.9$)& ($<5.0$)&  0.03  &   7.9 &  5.66 &  0.05 & (iii) \\
80 & 190229 & 7664 & 13102 & 3.46 &  10 &    ($<1.1$)& ($<4.8$) &  0.12  &   1.5 &  4.84 &  0.17 & (iii) \\
81 & 149121 & 6158 & 10748 & 3.89 &  10 &    ($<1.0$)& ($<5.2$) &  0.01  &   0.7 &  4.60 &  0.02 & (iii) \\
82 & 078316 & 3623 & 13513 & 3.85 &   8 &    ($<0.9$)& ($<4.8$) &  0.11  &   0.5 &  4.40 &  0.15 & (iii) \\
83 & 089822 & 4072 & 10307 & 3.89 &   5 &    ($<0.4$)& ($<5.0$) &  0.00  &   1.7 &  5.07 &  0.02 & (iii) \\
84 & 143807 & 5971 & 10828 & 4.06 &   4 &     3.3 &  5.82  &  0.01  &   9.0 &  5.85 &  0.02 & \\
85 & 193452 & 7775 & 10543 & 4.15 &   3 &     2.9 &  5.89  &  0.01  & \multicolumn{3}{c}{(no obs. data)} & \\
\hline
\end{tabular} 
\end{center}
(1) Star number (tentatively assigned). (2) Henry Draper Catalogue number. (3) Bright Star Catalogue number.
(4) Effective temperature (in K). (5) Logarithmic surface gravity (in cm\,s$^{-2}$/dex).
(6) Projected rotational velocity (in km\,s$^{-1}$). (7) Equivalent width of Al\,{\sc ii} 6243 line
(in m\AA). (8) Non-LTE Al abundance derived from Al\,{\sc ii} 6243 (in dex). (9) Non-LTE correction 
for Al\,{\sc ii} 6243 (in dex). (10) Equivalent width of Al\,{\sc ii} 4663 line (in m\AA). 
(11) Non-LTE Al abundance derived from Al\,{\sc ii} 4663 (in dex). (12) Non-LTE correction 
for Al\,{\sc ii} 4663 (in dex). (13) Remarks for the unmeasurable cases of Al\,{\sc ii} 6243.\\
Since this table is so arranged as to be consistent with Table~1 of Paper~II
(which should be consulted for more details of the program stars),
the first part (\#1--\#64) present the data of 64 early-to-late B stars 
(in the descending order of $T_{\rm eff}$) followed by the second part 
(\#65--\#85) for 21 late B-type stars (in the descending order of $v_{\rm e}\sin i$).
Column~(13)indicates the cases where Al abundances could not be established from the 
Al\,{\sc ii} 6243 line, which are divided into three types (cf. Sect.~4.3 and 4.4): 
(i) Line profiles show an appreciable emission feature (early-to-mid B-type stars). 
(ii) Lines are too weak because of filled-in emissions (early-to-mid B-type stars). 
(iii) Lines are too weak but considered to be due to very low Al abundance 
(late B-type HgMn stars). 
\end{table}

\setcounter{table}{2}
\begin{table}[H]
\scriptsize
\caption{Public-open data additionally employed in this study.}
\begin{center}
\begin{tabular}{cclll}\hline\hline
Star No. & HD\# & $^{*}$Data/Instrument & File name & Remark \\
\hline
 2 &  000886  &   ELODIE         &   \verb|elodie_19981123_0019.fits|         &  \\ 
 3 &  035708  &   CFHT/ESPaDOnS  &   \verb|1288837i.fits-1288844i.fits|       &  8 files co-added \\
 5 &  042690  &   ESO/FERROS     &   \verb|ADP.2016-09-27T09_50_43.972.fits|  &  \\
 7 &  034447  &   ESO/XSHOOTER   &   \verb|ADP.2017-08-11T08_26_08.731.fits|  &  \\
 9 &  043157  &   ESO/XSHOOTER   &   \verb|ADP.2017-08-10T15_20_15.958.fits|  &  $R= 3250$ \\
10 &  160762  &   CFHT/ESPaDOnS  &   \verb|1216549i.fits-1216568i.fits|       &  20 files co-added \\
14 &  025558  &   CFHT/ESPaDOnS  &   \verb|1977059i.fits-1977070i.fits|       & 12 files co-added \\
15 &  044700  &   ESO/XSHOOTER   &   \verb|ADP.2017-08-11T21_07_09.955.fits|  &  \\
17 &  181858  &   ESO/XSHOOTER   &   \verb|ADP.2017-08-18T07_38_30.899.fits|  &  $R=5400$ \\
19 &  185330  &   CFHT/ESPaDOnS  &   \verb|1266787i.fits-1266790i.fits|       &  4 files co-added \\
21 &  034798  &   ESO/FERROS     &   \verb|ADP.2016-09-27T09_50_43.812.fits|  & \\
23 &  198820  &   ELODIE         &   \verb|elodie_20041110_0019.fits|         & \\
26 &  037971  &   ELODIE         &   \verb|elodie_20030113_0020.fits|         & \\
27 &  026739  &   ESO/FERROS     &   \verb|ADP.2016-09-23T06_51_12.898.fits|  & \\
28 &  209008  &   ESO/UVES       &   \verb|ADP.2020-07-20T14_03_10.973.fits|  & \\
29 &  028375  &   ESO/XSHOOTER   &   \verb|ADP.2017-08-11T22_53_23.592.fits|  &  $R=5400$ \\
30 &  011415  &   CFHT/ESPaDOnS  &   \verb|1168487i.fits-1168490i.fits|       &  4 files co-added \\
33 &  189944  &   ESO/XSHOOTER   &   \verb|ADP.2017-08-11T18_51_15.004.fits|  & \\
34 &  181558  &   ESO/HARPS      &   \verb|ADP.2014-09-17T11_21_34.070.fits|  & \\
35 &  224990  &   ESO/FERROS     &   \verb|ADP.2016-09-27T09_50_42.585.fits|  & \\
36 &  175156  &   ESO/UVES       &   \verb|ADP.2020-06-30T13_28_27.894.fits|  & \\
39 &  202753  &   ESO/HARPS      &   \verb|ADP.2014-09-23T11_05_18.230.fits|  & \\
41 &  182255  &   ELODIE         &   \verb|elodie_20031106_0018.fits|         & \\
42 &  041692  &   ESO/HARPS      &   \verb|ADP.2016-09-05T01_02_13.863.fits|  & \\
43 &  049606  &   CFHT/ESPaDOnS  &   \verb|2787264i.fits-2787267i.fits|       &  4 files co-added \\
47 &  206540  &   ESO/XSHOOTER   &   \verb|ADP.2017-08-11T04_40_12.486.fits|  & \\
49 &  210424  &   ESO/UVES       &   \verb|ADP.2021-09-01T05_49_03.769.fits|  & \\
52 &  053244  &   ESO/FERROS     &   \verb|ADP.2016-09-21T07_46_52.316.fits|  & \\
53 &  155763  &   ELODIE         &   \verb|elodie_20010809_0017.fits|         &\\
54 &  173117  &   ESO/XSHOOTER   &   \verb|ADP.2018-09-13T17_16_55.500.fits|  &  $R=5400$ \\
55 &  017081  &   ESO/UVES       &   \verb|ADP.2020-07-10T21_52_27.364.fits|  & \\
56 &  023408  &   ESO/FERROS     &   \verb|ADP.2016-09-27T09_50_43.630.fits|  & \\
57 &  196426  &   ESO/UVES       &   \verb|ADP.2020-08-04T20_45_41.082.fits|  & \\
58 &  179761  &   ESO/UVES       &   \verb|ADP.2020-08-04T15_46_04.184.fits|  & \\
60 &  178065  &   ESO/UVES       &   \verb|ADP.2020-08-14T10_13_40.490.fits|  & \\
61 &  038899  &   ESO/HARPS      &   \verb|ADP.2016-09-04T01_02_02.711.fits|  & \\
62 &  043247  &   ESO/FERROS     &   \verb|ADP.2016-09-27T09_50_43.636.fits|  & \\
63 &  209459  &   CFHT/ESPaDOnS  &   \verb|1649409i.fits-1649412i.fits|       &  4 files co-added \\
65 &  098664  &   ESO/FERROS     &   \verb|ADP.2016-09-27T09_50_43.696.fits|  & \\
66 &  130557  &   ESO/UVES       &   \verb|ADP.2020-06-09T07_11_28.785.fits|  & \\
69 &  150100  &   ELODIE         &   \verb|elodie_19980704_0017.fits|         & \\
72 &  202671  &   ESO/FERROS     &   \verb|ADP.2016-09-27T09_50_43.826.fits|  & \\
73 &  193432  &   ESO/UVES       &   \verb|ADP.2020-08-14T10_35_52.467.fits|  & \\
75 &  077350  &   ELODIE         &   \verb|elodie_20050203_0013.fits|         & \\
78 &  144206  &   ELODIE         &   \verb|elodie_20000818_0006.fits|         & \\
79 &  145389  &   ELODIE         &   \verb|elodie_20040512_0019.fits|         & \\
80 &  190229  &   CFHT/ESPaDOnS  &   \verb|2770404i.fits-2770407i.fits|       &  4 files co-added \\
81 &  149121  &   ESO/UVES       &   \verb|ADP.2020-06-12T15_23_23.451.fits|  & \\
82 &  078316  &   ESO/FERROS     &   \verb|ADP.2016-09-27T07_02_43.215.fits|  & \\
83 &  089822  &   CFHT/ESPaDOnS  &   \verb|2948464i.fits-2948467i.fits|       &  4 files co-added \\
84 &  143807  &   ELODIE         &   \verb|elodie_20040510_0015.fits|         & \\
\hline
\end{tabular} 
\end{center}
These are the data adopted for the analysis of the Al\,{\sc ii} 4663 line (also for 
checking the Al\,{\sc ii} 3900 line in HD~160762, HD~209008, and HD~209459; cf. Sect.~5.2.3).
Although most of these spectra are of sufficiently high resolving power ($R \ga 10000$),
some are of medium spectral resolution ($R$ is only several thousands) as remarked 
in the last column.\\
ESO $\cdots$ ESO Science Archive Facility (\verb|https://archive.eso.org/cms.html|).\\
CFHT $\cdots$ Canada-France-Hawaii Telescope; data available from the Canadian Astronomy 
Data Centre (\verb|https://www.cadc.hia.nrc.gc.ca/|).\\
ELODIE $\cdots$ The ELODIE Archive (\verb|https://atlas.obs-hp.fr/elodie/|).\\
\end{table}

\end{document}